\documentclass[10pt]{article}

\usepackage{graphicx}
\usepackage{amssymb,amsfonts,amsmath}

\let\mpar=\marginpar
\renewcommand\marginpar[1]{\mpar{\raggedright \scriptsize #1}}

\begin{document}


\title{A First Exposure to Statistical Mechanics for Life Scientists}
\author{ Hernan G. Garcia$^{1}$, Jan\'e Kondev$^{2}$, Nigel Orme$^{3}$,
Julie A. Theriot$^{4}$, Rob Phillips$^{5}$
\\
\normalsize{$^{1}$Department of Physics, California Institute of
Technology, Pasadena, CA 91125, USA}\\
\normalsize{$^{2}$Department of Physics, Brandeis University
Waltham, MA 02454, USA}\\
\normalsize{$^{3}$Garland Science Publishing, 270 Madison Avenue, New York, NY 10016, USA}\\
\normalsize{$^{4}$Department of Biochemistry, Stanford
University School of Medicine, Stanford, CA 94305, USA}\\
\normalsize{$^{5}$Department of Applied Physics, California Institute of
Technology, Pasadena, CA 91125, USA}\\
}

\maketitle


\begin{abstract}

Statistical mechanics is one of the most powerful and elegant tools
in the quantitative sciences.      One key virtue of statistical
mechanics is that it is designed to examine large systems with many
interacting degrees of freedom, providing a clue that  it might have
some bearing on the analysis of the molecules of living matter.  As
a result of data on biological systems becoming increasingly
quantitative, there is a concomitant demand that the models set
forth to describe biological systems be themselves quantitative. We
describe how statistical mechanics is part of the quantitative
toolkit that is needed to respond to such data. The power of
statistical mechanics is not limited to traditional physical and
chemical problems and there are a host of interesting ways in which
these ideas can be applied in biology. This article reports on our
efforts to teach statistical mechanics to life science students and
provides a framework for others interested in bringing these tools
to a nontraditional audience in the life sciences.

\end{abstract}


\section{Does Statistical Mechanics Matter in Biology?}


The use of the ideas of equilibrium thermodynamics and statistical
mechanics to study biological systems are nearly as old as these
disciplines themselves.  Whether thinking about the binding
constants of transcription factors for their target DNA or proteins
on HIV virions for their target cell receptors, often the first
discussion of a given problem involves  a hidden assumption of
equilibrium. There are two key imperatives for students of the life
sciences who wish to explore the quantitative underpinnings of their
discipline: i) to have a sense of when the equilibrium perspective
is a reasonable approximation and ii) given those cases when it is
reasonable, to know how to use the key tools of the calculus of
equilibrium.  Our experiences in teaching both undergraduate and
graduate students in the life sciences as well as in participating
both as students and instructors in the Physiology Course at the
Marine Biological Laboratory in Woods Hole drive home the need for a
useful introduction to statistical mechanics for life scientists.


This paper is a reflection of our attempts to find a minimalistic
way of introducing statistical mechanics in the biological setting
that starts attacking biological problems that students might care
about as early as possible.   We view this as part of a growing list
of examples where quantitative approaches are included in the life
sciences curriculum \cite{Bio2010,Bialek2004,Wingreen2006}. As will
be seen throughout the paper, one of the key components of this
approach is to develop cartoons that provide a linkage between
familiar biological concepts and their mathematical incarnation. The
courses we teach  often involve a very diverse mixture of students
interested in how quantitative approaches from physics might be
useful for thinking about living matter.   On the one hand, we have
biology students that want to make the investment to learn tools
from physics.  At the same time, about one third of our students are
from that ever-growing category of physics students who are excited
about taking what they know and using it to study living organisms.
As a result, we face a constant struggle to not lose either the
interest or understanding of one of these two constituencies. The
challenge is to be interdisciplinary while maintaining strong
contact with the core disciplines themselves.

One of the key questions that must be addressed at the outset has to
do with the question of when equilibrium ideas are a viable approach
in thinking about real biological problems. Indeed, given the fact
that biological systems are in a constant state of flux, it is
reasonable to wonder whether equilibrium ideas are ever applicable.
Nevertheless,  there are a surprisingly large number of instances
when the time scales conspire to make the equilibrium approach a
reasonable starting point, even for examining some processes in
living cells.   To that end, we argue that the legitimacy of the
equilibrium approach often centers on the question of {\it relative}
time scales.    To be concrete, consider several reactions linked
together in a chain such as
\begin{equation}
\label{eq:DynamicEquilibrium}
 A \underset{k_-} {\overset{k_+}{\rightleftarrows}}B {\overset{r}{\rightarrow}}
 C \ .
\end{equation}
For simplicity, we consider a set of reactions in which the terminal
reaction is nearly irreversible. The thrust of our argument is that
even though the conversion of $B$ to $C$ is bleeding off material
from the $A$ and $B$ reaction, if the rate of $B$ to $C$ conversion
is sufficiently slow compared to the back reaction $B \rightarrow
A$, then the $A \underset{k_-} {\overset{k_+}{\rightleftarrows}}B$
reaction will always behave instantaneously as though it is in
equilibrium. There are a range of similar examples that illustrate
the way in which important biological problems, when boxed off
appropriately, can be treated from the equilibrium perspective
\cite{KeenerSneyd1998}. The goal of this paper is to use simple
model problems to illustrate how equilibrium ideas can be exploited
to examine biologically interesting case studies.


\section{Boltzmann, Gibbs and the Calculus of Equilibrium}
\label{Boltzmann}

We find that  statistical mechanics can be introduced in a
streamlined fashion by proceeding axiomatically.  We start  by
introducing a few key definitions and then arguing that just as
classical mechanics can be built exclusively around repeated uses of
${\bf F}=m{\bf a}$, statistical mechanics has its own fundamental
law (the Boltzmann distribution) from which results flow almost
effortlessly and  seemingly endlessly. There is a great deal of
precedent for this axiomatic approach as evidenced by several
amusing comments from well known statistical mechanics texts.  In
the preface to his book \cite{Mattis2003}, Daniel Mattis comments on
his thinking about what classes to take on statistical mechanics
upon his arrival at graduate school. ``I asked my classmate JR
Schrieffer, who presciently had enrolled in that class, whether I
should chance it later with a different instructor.  He said not to
bother - that he could explain all I needed to know about this topic
over lunch.  On a paper napkin, Bob wrote $e^{-\beta H}$ ``That's it
in a nutshell''. ``Surely you must be kidding Mr. Schrieffer'' I
replied (or words to that effect)  ``How could you get the
Fermi-Dirac distribution out of THAT?''  ``Easy as pie'' was the
reply ... and I was hooked".

Similarly, in speaking of the Boltzmann distribution,  Feynman notes
in the opening volley of his statistical mechanics book: ``This
fundamental law is the summit of statistical mechanics, and the
entire subject is either the slide-down from this summit, as the
principle is applied to various cases, or the climb-up to where the
fundamental law is derived...'' \cite{Feynman3}. Our sense is that
in a first exposure to statistical mechanics for students of the
life sciences, an emphasis on the slide-down from the summit which
illustrates the intriguing applications of statistical mechanics is
of much greater use than paining through the climb to that summit
with a consideration of the nuances associated with where these
distributions come from. As a result, we relegate a derivation of
the Boltzmann distribution to the appendix at the end of this paper.


So what is this ``summit'' that Feynman speaks of?    Complex,
many-particle systems such as the template DNA, nucleotides, primers
and enzymes that make up a polymerase chain reaction, familiar to
every biology student, can exist in an astronomical number of
different states.
It is the job of statistical mechanics to assign probabilities to
all of these different ways (the distinct ``microstates'') of
arranging the system.    The summit that Feynman speaks of is the
simple idea that each of these different arrangements has a
probability proportional to $e^{-\beta E_i}$, where $E_i$ is the
energy of the microstate of interest which is labeled by the index
$i$.    To make this seem less abstract, we begin our analysis of
statistical mechanics  by describing the notion of a microstate in a
way that will seem familiar to biologists.

\begin{figure}
\begin{center}
\includegraphics[width=5.5in]{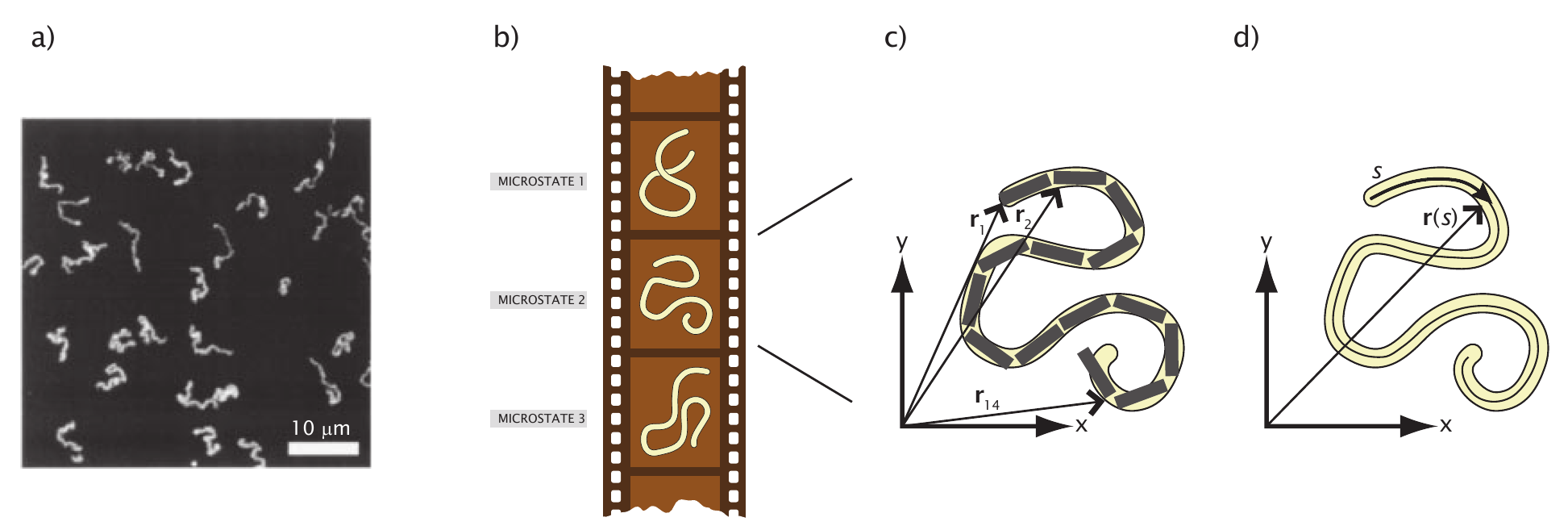}
\caption{Microstates of DNA in solution.  (a) Fluorescence microscopy images
of $\lambda$-phage DNA \cite{Maier1999} (reprinted with permission). The DNA molecule jiggles
around in solution and every configuration corresponds to a different
microstate. (b)  The film strip shows how at every instant at which a picture is
taken, the DNA configuration is different.   From a mathematical
perspective, we can represent the configuration of the molecule either by
using  (c) a discrete set of vectors $\{{\bf r}_i\}$ or  (d) by the continuous
function  ${\bf r}(s)$.
}
\label{fig:Microstates}
\end{center}
\end{figure}

One of our favorite examples for introducing the concept of a
microstate is to consider a piece of DNA from the bacterial virus
known as $\lambda$-phage. If one of these $\approx$ 48,500~base pair
long DNA molecules is fluorescently labeled and observed through a
microscope as it jiggles around in solution, we argue that the
different conformations adopted by the molecule correspond to its
different allowed microstates. Of course, for this idea to be more
than just words, we have to invoke some mathematical way to
represent these different microstates.   As shown in
fig.~\ref{fig:Microstates}, it is possible to characterize the
states of a polymer such as DNA either discretely (by providing the
$x,y,z$ coordinates of a set of discrete points on the polymer)  or
continuously (by providing the position of each point on the
polymer, ${\bf r}(s)$, as a function of the distance $s$ along the
polymer).

A second way in which the notion of a microstate can be introduced
that is biologically familiar  is by discussing ligand-receptor
interactions (and binding interactions more generally).   This topic
is immensely important in biology as can be seen  by appealing to
problems such as antibody-antigen binding, ligand-gated ion channels
and oxygen binding to hemoglobin, for example.  Indeed, Paul Ehrlich
made his views on the importance of ligand-receptor binding evident
through the precept: ``Corpora non agunt nisi ligata -  A substance
is not effective unless it is linked to another'' \cite{Klotz1997}.
Whether discussing signaling, gene regulation or metabolism,
biological action is a concert of different binding reactions and we
view an introduction to the biological uses of binding and how to
think about such binding using statistical mechanics as a worthy
investment for biology and physics students alike.

\begin{figure}
\begin{center}
\includegraphics[width=2.5in]{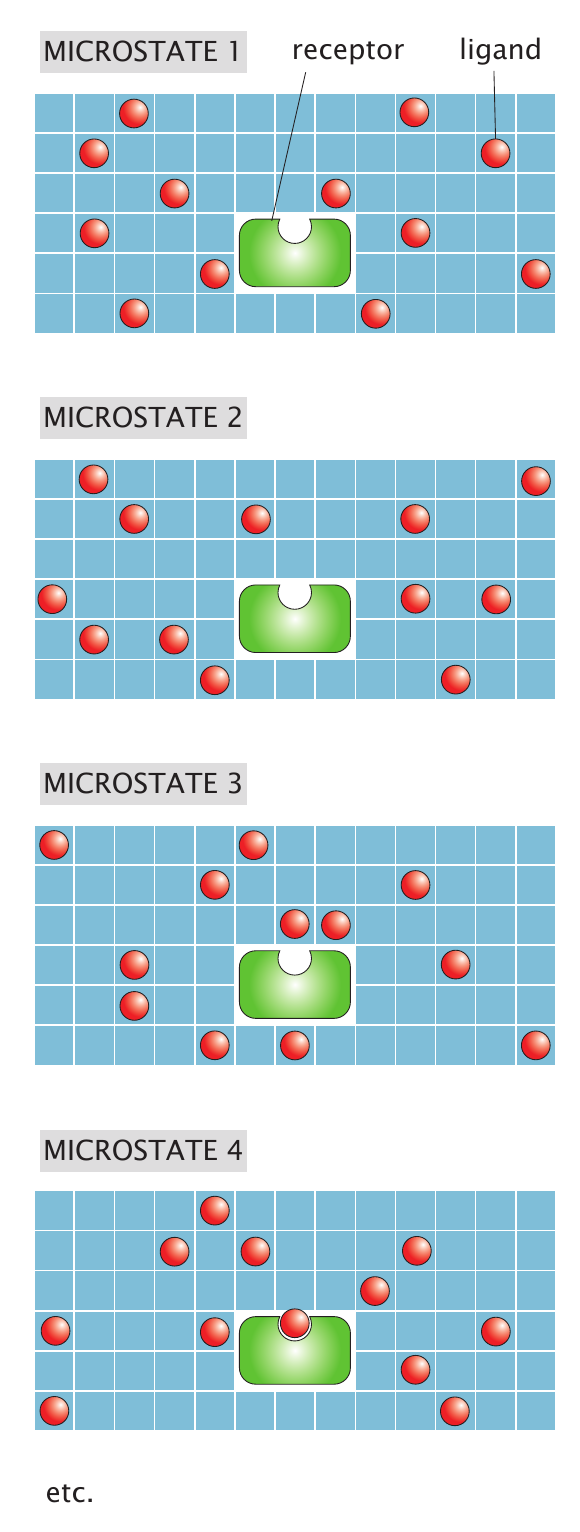}
\caption{Lattice model for solution.  Ligands in solution with their
partner receptor. A simplified lattice model posits a discrete set
of sites (represented as boxes)  that the ligands can occupy and
permits a direct evaluation of the various microstates (ways of
arranging the ligands).  The first three microstates shown here
have the receptor unoccupied while the fourth microstate is one
in which the receptor is occupied.} \label{LatticeModel}
\end{center}
\end{figure}

To treat cases like these, it is
 convenient to imagine an isolated system represented by a
box of solution which contains a single receptor and $L$ ligands.
One of the pleasures of using this example is that it emphasizes
 the simplifications that physicists love \cite{Dill2003}.
In particular, as shown in fig.~\ref{LatticeModel}, we introduce a
lattice model of the solution in which the box is divided up into
$\Omega$ sub-boxes. These sub-boxes have molecular dimensions and
can be occupied by only one ligand molecule at a time. We also
assume that the concentration of ligand is so low that they do not
interact. In this case, the different microstates correspond to the
different ways of arranging the $L$ ligands amongst these $\Omega$
elementary boxes. Although it is natural for biological scientists
who are accustomed to considering continuous functions and
concentrations to chafe against the discretization of a solution in
a lattice model, it is fairly easy to justify this simplification.
We may choose any number of boxes $\Omega$, of any size.  At the
limit of a large number of very small boxes, they may have molecular
dimensions.  In practice, the mathematical results are essentially
the same for most choices where $\Omega \gg L$, that is, where the
solution is dilute. Given $L$ ligands and $\Omega$ sites that they
can be distributed on, the total number of microstates available to
the system (when no ligands are bound to the receptor) is
\begin{equation}
    \mbox{number of microstates} = {\Omega! \over L! (\Omega -L)!}.
\label{CountingDegeneracy}
\end{equation}
The way to see this result is to notice that for the first ligand,
we have $\Omega$ distinct possible places that we can put the
ligand. For the second ligand, we have only $\Omega-1$ choices and
so on. However, once we have placed those ligands, we can
interchange them in $L!$ different ways without changing the actual
microstate.  In fact, these issues of rearrangement
(distinguishability vs. indistinguishability) are subtle, but
unimportant for our purposes since they don't change the ultimate
results for classical systems such as the biological problems
addressed here. We leave it as an exercise for the reader to show
that the results in this paper would remain unaltered if considering
the ligands to be distinguishable.

During our participation in the MBL Physiology Course in the summer
of 2006, our favorite presentation of this notion of microstate was
introduced by a PhD biology student from  Rockefeller University.
During her presentation, she used a box from which she cut out a
square in the bottom.  She then taped a transparency onto the bottom
of the box and drew a square grid precisely like the one  shown in
fig.~\ref{LatticeModel} and then constructed a bunch of
``molecules'' that just fit into the squares.  Finally, she put this
box on an overhead projector and shook it, repeatedly demonstrating
the different microstates available to ``ligands in solution''.

These different examples of microstates also permit us to give a cursory introduction
to the statistical mechanical definition of entropy.   In particular, we introduce entropy as
a measure of microscopic degeneracy through the expression
\begin{equation}
    S(V,N)=k_B \mbox{ln}  ~W(V,N),
\end{equation}
where $k_{B} = 1.38 \times 10^{-23}~\rm{J/K}$ the all-important
constant of statistical mechanics known as the Boltzmann constant,
$S(V,N)$ is the entropy of an isolated system with volume $V$
containing $N$ particles, and $W(V,N)$ is the number of distinct
microstates available to that system.      We also argue that the
existence of the entropy function permits the introduction of the
all-important variational statement of the second law of
thermodynamics which tells us how to select out of all of the
possible macrostates of a system, which state is most likely to be
observed. In particular, for an isolated system (i.e. one that has
rigid, adiabatic, impermeable walls, where no matter or energy can
exit or enter the system) the equilibrium state is that macrostate
that maximizes the entropy. Stated differently, the macroscopically
observed state will be that state which can be realized in the
largest number of microscopic states.

Now that we have the intuitive idea of microstates in hand {\it and}
have shown how to enumerate them mathematically, and furthermore we
have introduced Gibbs' calculus of equilibrium in the form of the
variational statement of the second law of thermodynamics, we are
prepared to introduce the Boltzmann distribution itself. The
Boltzmann distribution derives naturally from the second law of
thermodynamics as the distribution that maximizes the entropy of a
system in contact with a thermal bath (for a detailed derivation
refer to the appendix). Statistical mechanics describes systems in
terms of the probabilities of the various microstates. This style of
reasoning is different from the familiar example offered by
classical physics in disciplines such as mechanics and electricity
and magnetism which centers on deterministic analysis of physical
systems. By way of contrast, the way we do ``physics" on systems
that can exist in astronomical numbers of different microstates is
to assign probabilities to them.

As noted above, one useful analogy is with the handling of
classical dynamics.  All science  students have at one time or another
been taught Newton's second law of motion (${\bf F}=m{\bf a}$) and usually, this
governing equation is introduced axiomatically.   There
is a corresponding
central equation in statistical mechanics which can also
be introduced axiomatically.  In particular,
if we label the $i^{th}$ microstate by its energy $E_i$, then the
probability of that microstate is given by
\begin{equation}
    p_{i}={1 \over Z} e^{-\beta E_i},
\end{equation}
where $\beta=1/k_BT$. The factor $e^{-\beta E_i}$ is called the
Boltzmann Factor and $Z$ is the ``partition function'', which is
defined as $Z= \sum_i e^{-\beta E_i}$. Note that from this
definition it follows that the probabilities are normalized, namely
$\sum_{i} p_{i} = 1$. Intuitively, what this distribution tells us
is that when we have a system that can exchange energy with its
environment, the probability of that system being in a particular
microstate decays exponentially with the energy of the microstates.
Further, $k_BT$ sets the natural energy scale of physical biology
where the temperature T for biological systems is usually around 300
K by telling us that microstates with energies too much larger than
$k_BT \approx 4.1 ~pN~nm \approx 0.6~kcal/mol \approx 2.5~kJ/mol$
are thermally inaccessible. This first introduction to the Boltzmann
distribution suffices to now begin to analyze problems of biological
relevance.



%


\section{State Variables and States and Weights}
\label{StatesWeightsSection}

%

\begin{figure}
\begin{center}
\includegraphics[width=3.3in]{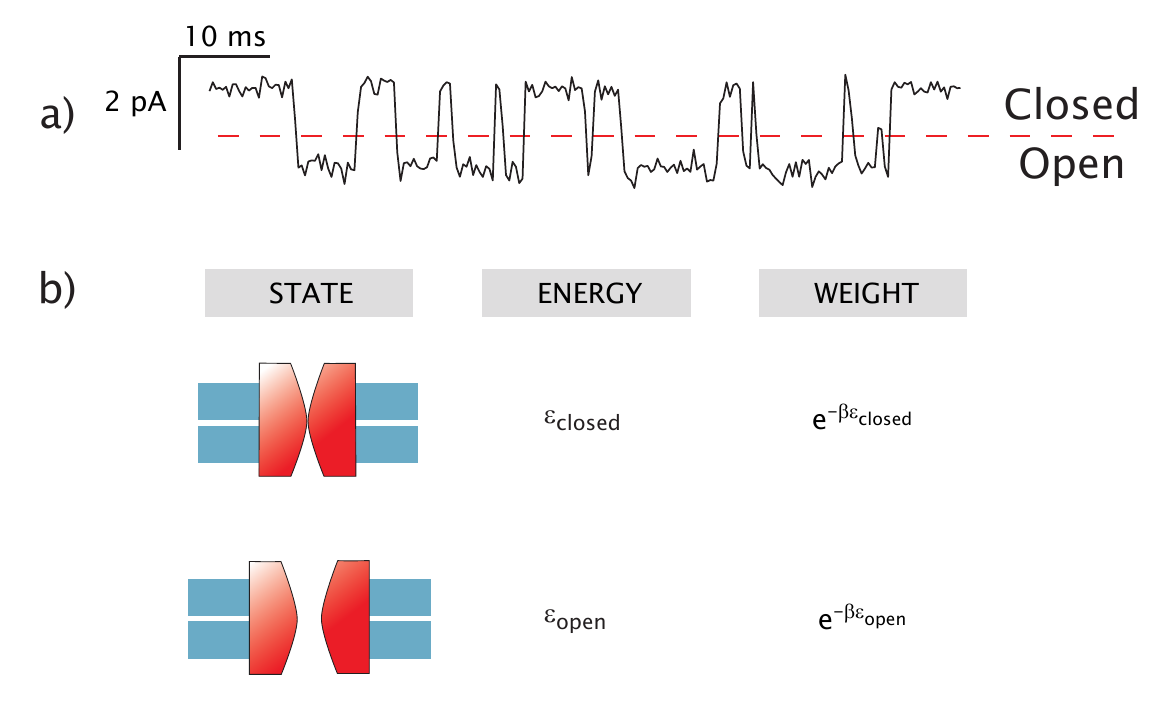}
\caption{Ion channel open probability.  (a) Current as a function of time
for an ion channel showing repeated transitions between
the open and closed states \cite{Grosman2000}.  (b)  States and weights for an ion channel.  The cartoon shows a
schematic of the channel states which have different energies, and
by the Boltzmann distribution, different probabilities.}
\label{StatesWeightsSchematic}
\end{center}
\end{figure}

One way to breathe  life into the Boltzmann distribution is by
constructing a compelling and honest correspondence between
biological cartoons and their  statistical mechanical meaning. Many
of the cartoons familiar from molecular biology textbooks are
extremely information-rich
 representations of a wealth  of biological data and understanding.
These informative cartoons can often be readily adapted to a
statistical mechanics analysis simply by assigning statistical
weights to the different configurations as shown in
fig.~\ref{StatesWeightsSchematic}.    This first example considers
the probability of finding an ion channel in the open or closed
state in a simple model in which it is assumed that the channel can
exist in only two states. We use ion channels as our first example
of states and weights because, in this way, we can appeal to one of
the physicists favorite models, the two-level system, while at the
same time using a biological example that is of central importance.
During our course, we try to repeat this same basic motif of
identifying in cartoon form the microscopic states of some
biological problem  and then to assign those different states and
their corresponding statistical weights (and probabilities).

As shown in fig.~\ref{StatesWeightsSchematic}, in the simplest model
of an ion channel we argue that there are two states: the closed
state and the open state. In reality some ion channels have been
shown to exist in more than two states. However, the two state
approximation is still useful for understanding most aspects of
their electrical conductance behavior as shown by the trace in
fig.~\ref{StatesWeightsSchematic}a where only two states
predominate. In order to make a mapping between the biological
cartoon and the statistical weights, we need to know the energies of
the closed and open states, $\epsilon_{\text{closed}}$ and
$\epsilon_{\text{open}}$. Also, for many kinds of channels the
difference in energy between closed and open can be tuned by the
application of external variables such as an electric field
(voltage-gated channels), membrane tension (mechanosensitive
channels) and the binding of ligands (ligand-gated channels)
\cite{Woolley2003}.

We find that a two-state ion channel is one of the cleanest and
simplest examples for introducing how a statistical mechanics
calculation might go. In addition, these ideas on ion channels serve
as motivation for the introduction of a convenient way of
characterizing the ``state'' of many macromolecules of biological
interest. We define the two-state variables $\sigma$ which can take
on either the values $0$ or $1$ to signify the distinct conformation
or state of binding of a given molecule. For example, in the ion
channel problem $\sigma=0$ corresponds to the closed state of the
channel and $\sigma=1$ corresponds to the open state (this choice is
arbitrary, we could equally have chosen to call $\sigma=1$ the
closed state, but this choice makes more intuitive sense).  As a
result, we can write the energy of the ion channel as
\begin{equation}
E(\sigma)=(1-\sigma)\epsilon_{\text{closed}}+\sigma
\epsilon_{\text{open}}.
\end{equation}
Thus, when the channel is closed, $\sigma=0$ and the energy is
$\epsilon_{\text{closed}}$. Similarly, when $\sigma=1$, the channel
is open and the energy is $\epsilon_{\text{open}}$. Though this
formalism may seem heavy handed, it is extremely convenient for
generalizing to more complicated problems such as channels that can
have a ligand bound or not as well being open or closed. Another
useful feature of this simple notation is that it permits us to
compute quantities of experimental interest straight away. Our aim
is to compute the open probability $P_{\text{open}}$ which, in terms
of our state variable $\sigma$, can be written as $\langle \sigma
\rangle$, where $\langle \cdots \rangle$ denotes an average. When
$\langle \sigma \rangle \approx 0$ this means that the probability
of finding the channel open is low. Similarly, when $\langle \sigma
\rangle \approx 1$, this means that it is almost certain that we
will find the channel open.  The simplest way to think of this
average is to imagine a total of $N$ channels and then to evaluate
the fraction of the channels that are in the open state,
$N_{open}/N$.


To compute the probability that the channel will be open, we invoke
the Boltzmann distribution and, in particular, we evaluate the
partition function given by
\begin{equation}\label{eq:ZOpenClose}
Z=\sum_{\sigma=0}^1 e^{-\beta E(\sigma)}=e^{-\beta
\epsilon_{\text{closed}}}+e^{-\beta \epsilon_{\text{open}}}.
\end{equation}
As noted above, for the simple two-state description of a channel,
the partition function is a sum over only two states, the closed
state and the open state. Given the partition function, we then know
the probability of both the open and closed states via,
$p_{open}=e^{-\beta \epsilon_{\text{open}}}/Z$ and
$p_{\text{closed}}=e^{-\beta \epsilon_{\text{closed}}}/Z$.

\begin{figure}
\begin{center}
\includegraphics[width=3.3in]{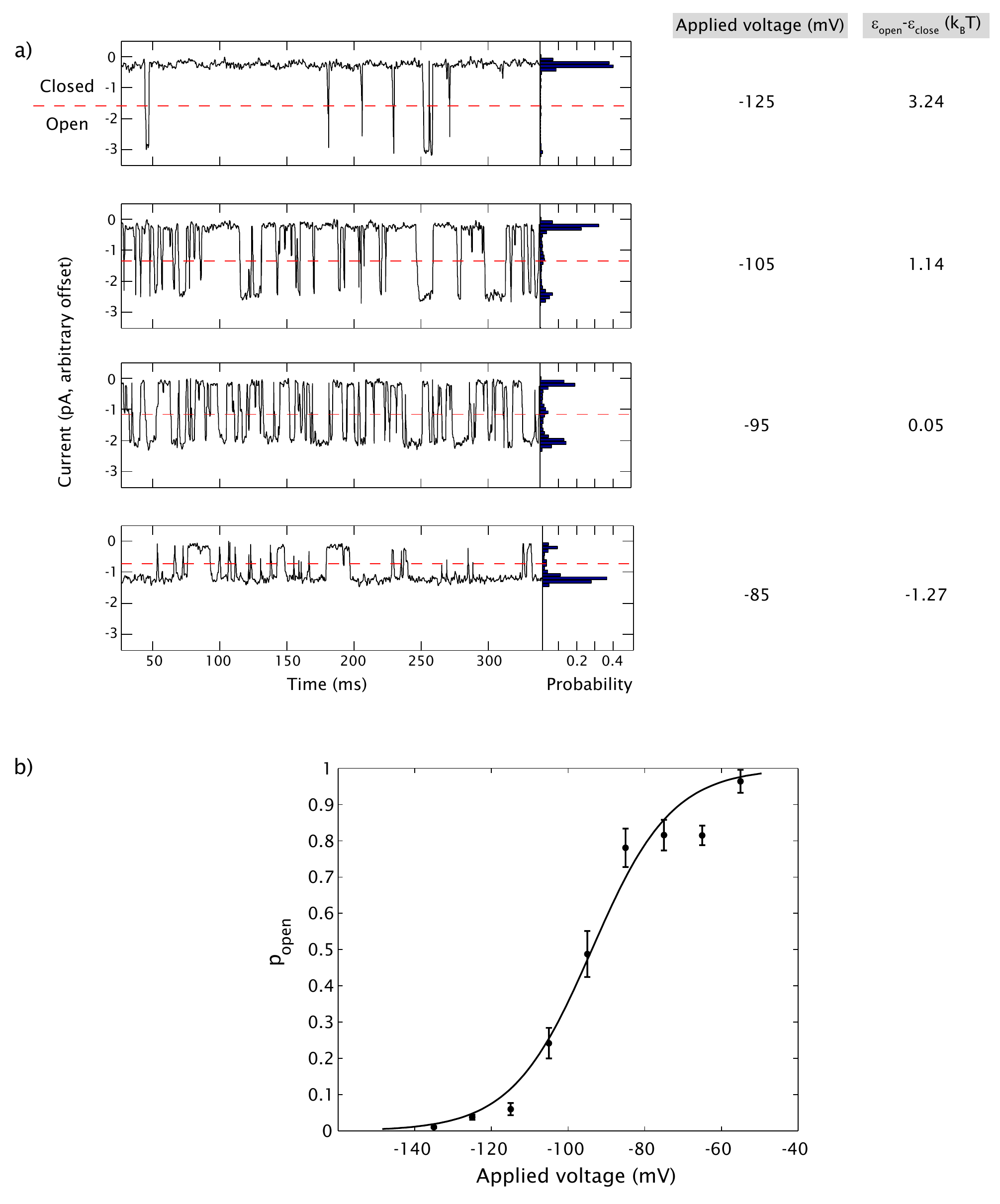}
\caption{Probability of open and closed states for a voltage-sensitive ion channel. (a) Current traces for
a sodium channel for four different applied voltages.   The histograms to the right of each current
trace show the fraction of time spent in the two states \cite{Keller1986}. (b)
Probability that the channel is open as a function of the applied voltage.  The data points
correspond to computing the fraction of time the channel spends open for traces like those shown
in (a).  The curve shows a fit to the data using
eq.~\ref{channelEquation}, where
$\beta (\epsilon_{\text{open}}-\epsilon_{\text{closed}}) = \beta \alpha (V_{\rm{applied}}-V_{0})$.
The fit yields $\alpha \simeq -0.096~k_{B}T/\text{mV}$ and $V_{0} = -94~\text{mV}$.}
\label{fig:ChannelProb}
\end{center}
\end{figure}

Using the partition function of eq.~\ref{eq:ZOpenClose}, we see that
the open probability is given by (really, it is nothing more than
$p_{\text{open}}$)
\begin{equation}
    \langle \sigma \rangle =\sum \sigma p(\sigma)= 0 \times p(0) + 1 \times p(1).
\end{equation}
As a result, we see that the open probability is given by
\begin{equation}
\langle \sigma \rangle ={e^{-\beta \epsilon_{\text{open}}} \over
e^{-\beta \epsilon_{\text{closed}}}+e^{-\beta
\epsilon_{\text{open}}}}=\frac{1}{e^{\beta(\epsilon_{\text{open}}-\epsilon_{\text{closed}})}+1}.
\label{channelEquation}
\end{equation}
This result illustrates more quantitatively the argument made
earlier that the energy scale $k_BT$ is the standard that determines
whether a given microstate is accessible. It also shows how in terms
of energy what really matters is the relative difference between the
different states ($\Delta \epsilon =
\epsilon_{\text{open}}-\epsilon_{\text{closed}}$) and not their
absolute values. An example of the probability of the channel being
open is shown in fig.~\ref{fig:ChannelProb} for several different
choices of applied voltage in the case where the voltage is used to
tune the difference between $\epsilon_{open}$ and
$\epsilon_{closed}$. In turn, measuring $p_{\rm{open}}$
experimentally tells you $\Delta \epsilon$ between the two states.

Often, when using statistical mechanics to analyze problems of
biological interest, our aim is to characterize several features of
a macromolecule at once.  For example, for an ion channel, the
microstates are described by several ``axes'' simultaneously. Is
there a bound ligand?  Is the channel open or not?  Similarly, when
thinking about post-translational modifications of proteins, we are
often interested in the state of phosphorylation of the protein of
interest \cite{Johnson2001,walsh}.     But at the same time, we
might wish to characterize the protein as being active or inactive,
and also whether or not it is bound to allosteric activators or
inhibitors. As with the ion channel, there are several variables
needed to specify the overall state of the macromolecule of
interest. Although this is admittedly oversimplified, countless
biologically relevant problems can be approached by considering two
main states (bound vs. unbound, active vs. inactive, phosphorylated
vs. unphosphorylated, etc.) each of which can be characterized by
its own state variable $\sigma_i$.   The key outcome of this first
case study is that we have seen how both the Boltzmann distribution
and the partition function are calculated in a simple example and
have shown a slick and very useful way of representing biochemical
states using simple two-state variables.

\section{Ligand-Receptor Binding as a Case Study}
\label{LigandReceptorExample}

The next step upward in complexity is to consider ligand-receptor
interactions where we must keep track of two or more separate
molecules rather than just one molecule in two different states.
Examples of this kind of binding include: the binding of
acetylcholine to the nicotinic acetylcholine receptor
\cite{Zhong1998}, the binding of transcription factors to DNA
\cite{Bintu2005a}, the binding of oxygen to hemoglobin, the binding
of antigens to antibodies \cite{Janeway2005} and so on. To examine
the physics of fig.~\ref{LatticeModel}, imagine there are $L$ ligand
molecules in the box characterized by $\Omega$ lattice sites as well
as a single receptor with one binding site as shown.    Our ambition
is to compute the probability that the receptor will be
occupied ($p_{bound}$) as a function of the number (or
concentration) of ligands.

\begin{figure}
\begin{center}
\includegraphics[height=2.0truein]{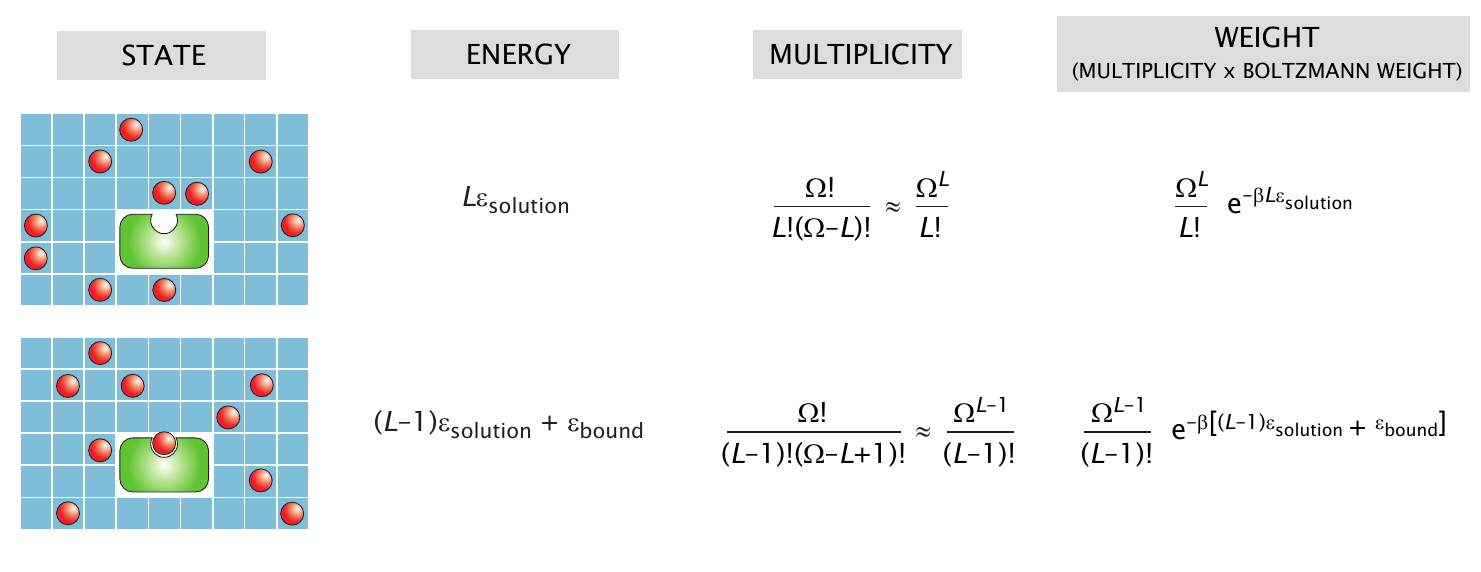}
\caption[]{States and weights diagram for ligand-receptor binding.
The cartoons show a lattice model of solution for the case in which
there are $L$ ligands and $\Omega$ lattice sites.  In the upper panel,
the receptor is unoccupied.  In the lower panel, the receptor is
occupied by a ligand and the remaining $L-1$ ligands are free in
solution.   A given state has a weight dictated by its Boltzmann
factor.    The multiplicity refers to the number of different
microstates that share that same Boltzmann factor (for example, all
of the states with no ligand bound to the receptor have the same
Boltzmann factor).  The total statistical weight is given by the
product of the multiplicity and the Boltzmann factor.}
\label{StatesWeightsReceptorLigand}
\end{center}
\end{figure}

To see the logic of this calculation more clearly,
fig.~\ref{StatesWeightsReceptorLigand} shows the states available to
this system as well as their Boltzmann factors, multiplicities (i.e.
the number of different ways of arranging $L$ or $L-1$  ligands in
solution) and overall statistical weights which are the products of
the multiplicities and the Boltzmann factor. The key point is that
there are only two classes of states: i) all of those states for
which there is no ligand bound to the receptor and ii) all of those
states for which one of the ligands is bound to the receptor.  The
useful  feature of this situation is that although there are many
realizations of each class of state, the Boltzmann factor for each
of these individual realizations for each of the classes of state
are all the same as shown in fig.~\ref{StatesWeightsReceptorLigand} since
all microstates in each class have the same energy.

\begin{figure}
\begin{center}
\includegraphics[width=4.5truein]{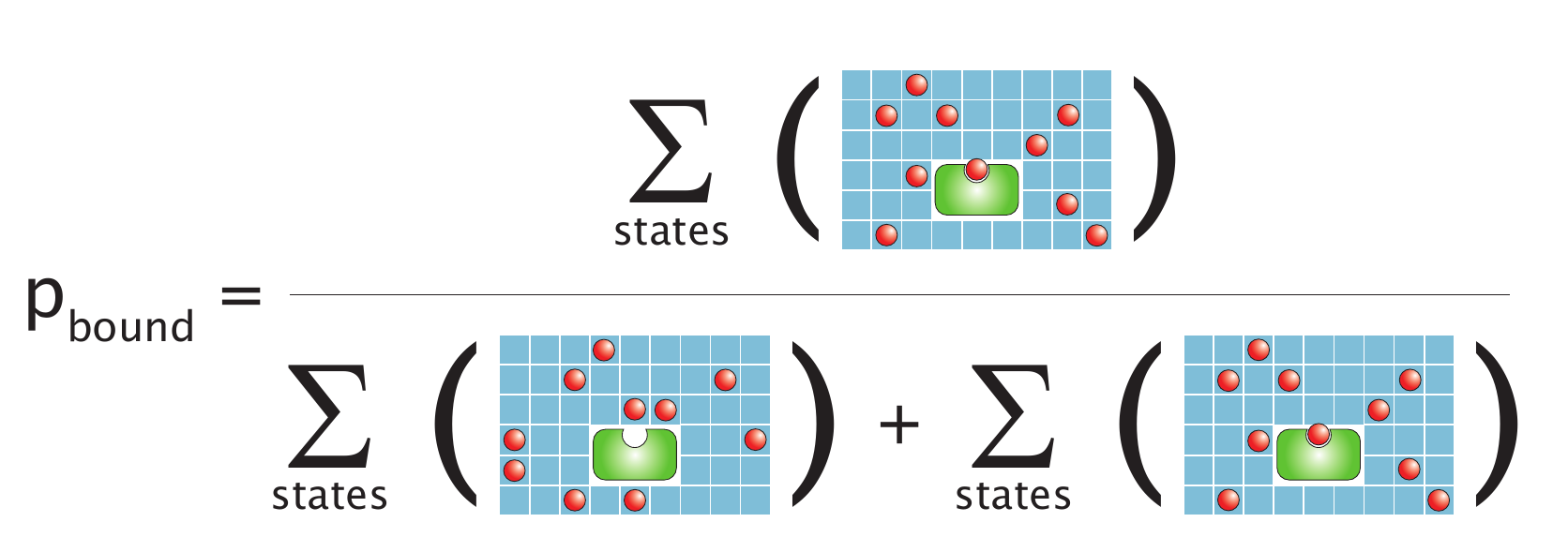}
\caption[]{Probability of receptor occupancy.  The figure shows how
the probability of receptor occupancy can be written as a ratio of
the weights of the favorable outcomes and the weights of {\it all}
outcomes.  In this case the numerator is the result of summing over the weights of
all states in which the receptor is occupied.} \label{pBoundReceptorLigand}
\end{center}
\end{figure}

To compute the probability that the receptor is occupied, we need to
construct a ratio in which the numerator involves the
accumulated statistical weight of  all states
in which one ligand is bound to the  receptor and the denominator is
the sum over all states.   This idea is represented graphically in
fig.~\ref{pBoundReceptorLigand}.   What the figure shows is that
there are a host of different states in which the receptor is
occupied: first, there are $L$ different ligands that can bind to
the receptor, second, the $L-1$ ligands that remain behind in
solution can be distributed amongst the $\Omega$ lattice sites in many
different ways.    In particular, we have
\begin{equation}
\mbox{weight when receptor occupied}=\underbrace{e^{-\beta
\epsilon_{\text{bound}}}}_{\mbox{receptor}} \times
\underbrace{\sum_{\mbox{solution}}e^{-\beta (L-1)
\epsilon_{\text{solution}}},}_{\mbox{free ligands}}
\end{equation}
where we have introduced $\epsilon_{\text{bound}}$ as the energy for
the ligand when bound to the receptor and
$\epsilon_{\text{solution}}$ as the energy for a ligand in solution.
The summation $\sum_{\mbox{solution}}$ is an instruction to sum over
all of the ways of arranging the $L-1$ ligands on the $\Omega$
lattice sites in solution with each of those states assigned the
weight $e^{-\beta (L-1) \epsilon_{\text{solution}}}$. Since the
Boltzmann factor is the same for each of these states, what this sum
amounts to is finding the number of arrangements of the $L-1$
ligands amongst the $\Omega$ lattice sites and yields
\begin{equation}
\sum_{\mbox{solution}}e^{-\beta (L-1)
\epsilon_{\text{solution}}}=e^{-\beta (L-1)
\epsilon_{\text{solution}}} {\Omega! \over (L-1)! (\Omega-(L-1))!}.
\label{LigandReceptorSum1}
\end{equation}
To effect this sum, we have exploited precisely the same counting
argument that led to eq.~\ref{CountingDegeneracy} with the only
change that now we have $L-1$ ligands rather than $L$. The
denominator of the expression shown in
fig.~\ref{pBoundReceptorLigand} is  the partition function itself
since it represents a sum over {\it all} possible arrangements of
the system (both those with the receptor occupied and not) and is
given by
\begin{equation}\label{eq:Zligandreceptor}
Z(L,\Omega)=\underbrace{\sum_{\mbox{solution}}e^{-\beta L
\epsilon_{\text{solution}}}}_{\mbox{none bound}} +
\underbrace{e^{-\beta \epsilon_{\text{bound}}}
\sum_{\mbox{solution}} e^{-\beta (L-1)
\epsilon_{\text{solution}}}}_{\mbox{ligand bound}}.
\end{equation}
We already evaluated the second term in the sum culminating in
eq.~\ref{LigandReceptorSum1}. To complete our evaluation of the
partition function,
 we have to evaluate the sum $\sum_{\mbox{solution}}e^{-\beta L \epsilon_{\text{solution}}}$
over all of the ways of arranging the $L$ ligands on the $\Omega$ lattice sites
with the result
\begin{equation}
\sum_{\mbox{solution}}e^{-\beta L
\epsilon_{\text{solution}}}=e^{-\beta L \epsilon_{\text{solution}}}
{\Omega! \over L! (\Omega-L)!}.
\end{equation}
In light of these results, the partition function  can be written as
\begin{equation}\label{eq:ZLigandReceptor}
Z(L,\Omega)=e^{-\beta L \epsilon_{\text{solution}}}
    \left[ {\Omega! \over L! (\Omega-L)!} \right] +
    e^{-\beta \epsilon_{\text{bound}}} e^{-\beta (L-1)
    \epsilon_{\text{solution}}} \left[ {\Omega! \over (L-1)! (\Omega-(L-1))!} \right].
\end{equation}
We can now simplify this result by using the approximation that
\begin{equation}
{\Omega! \over (\Omega-L)!} \approx \Omega^L,
\end{equation}
which is justified as long as $\Omega >> L$.  To see why this is a good
approximation consider the case when  $\Omega = 10^{6}$ and $L=10$ resulting in
\begin{equation}
    \frac{10^{6}!}{(10^{6} - 10)!} = 10^{6} \cdot (10^{6} - 1) \cdot (10^{6} -2) \cdot ... \cdot (10^{6} -9) \simeq (10^{6})^{10}.
\end{equation}
Note that the approximate result is the largest term in the sum
which we obtain by multiplying out all the terms in parentheses. We
leave it to the reader to show that the correction is of the order
$0.001~\%$, five orders of magnitude smaller than the leading term.

%

\begin{figure}
\begin{center}
\includegraphics[height=3.0truein]{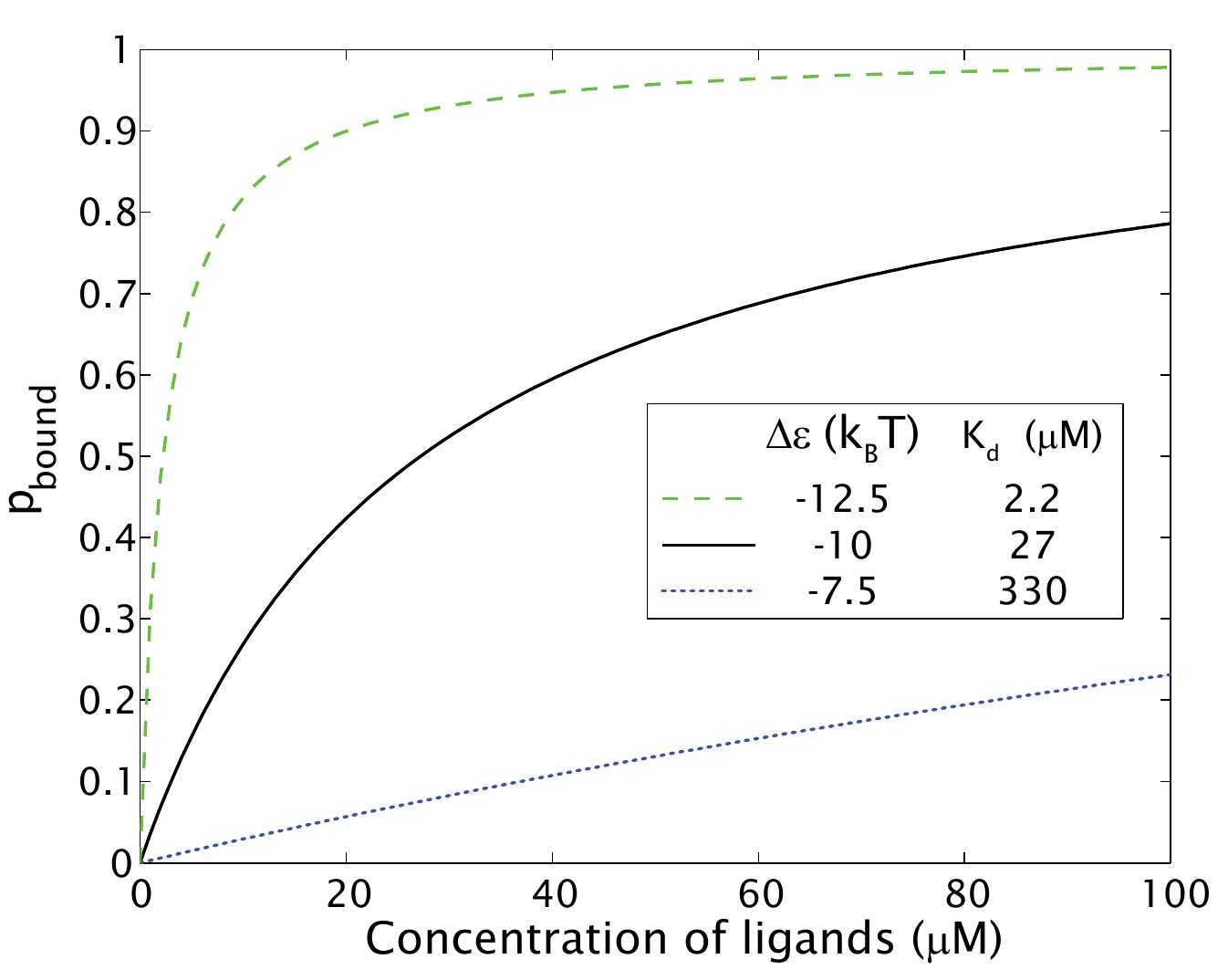}
\caption[]{Simple binding curve.  This curve shows
$p_{\text{bound}}$ as calculated in
eq.~\ref{pBoundLigandReceptorEquation}.  The different curves
correspond to different choices for the strength of the
ligand-receptor binding energy, $\Delta \epsilon$, given a reference
concentration of $c_{0} = 0.6~\rm{M}$. The corresponding dissociation
constants are shown as well.}
\label{BindingCurve}
\end{center}
\end{figure}

\begin{figure}
\begin{center}
\includegraphics[width=3.3in]{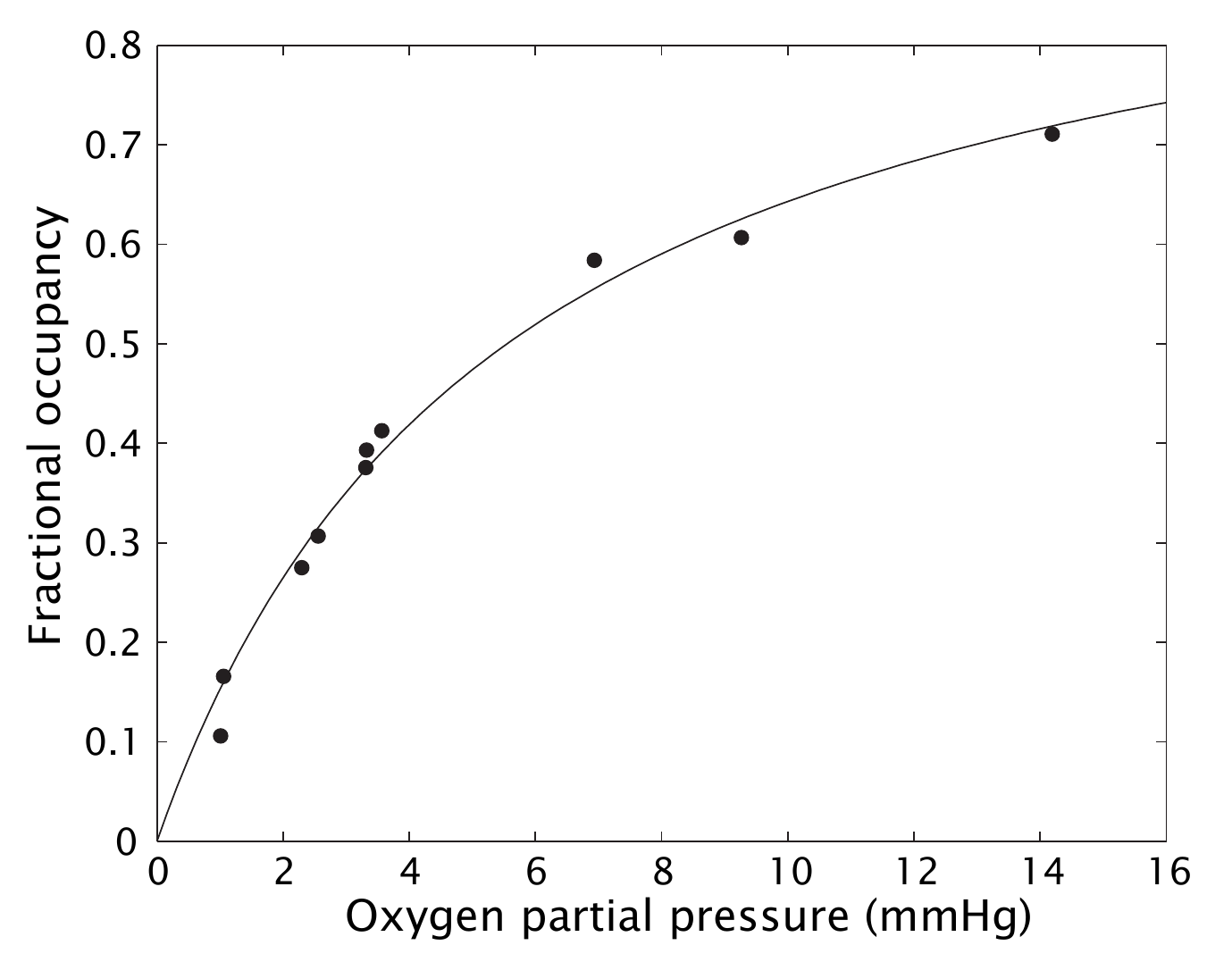}
\caption{Example of ligand-receptor binding.   The binding of oxygen to myoglobin
as a function of the oxygen partial pressure.  The points correspond to
the measured occupancy of myoglobin as a function of the oxygen partial pressure \cite{Changeux1965}
(because oxygen is a gas, partial pressure is used rather than concentration)
and the curve is a fit based upon the one-parameter model from eq.~\ref{pBoundLigandReceptorEquation}.
The fit yields $\Delta \epsilon = -4.9 k_{B}T$ using a standard state $c_0 = 760~\text{mmHg} = 1~\text{atm}$,
which also corresponds to a dissociation constant $K_{d} = 5.6~\text{mmHg}$.}
\label{fig:MyoBound}
\end{center}
\end{figure}

%
 With these results  in hand, we can now write $p_{bound}$ as
\begin{equation}
p_{bound}=\frac{e^{-\beta \epsilon_{\text{bound}}}
    \frac{\Omega^{L-1}}{(L-1)!} e^{-\beta (L-1)
    \epsilon_{\text{solution}}}}
    {\frac{\Omega^L}{L!} e^{-\beta
    L\epsilon_{\text{solution}}}
    +e^{-\beta \epsilon_{\text{bound}}}
    \frac{\Omega^{L-1}}{(L-1)!} e^{-\beta (L-1)
    \epsilon_{\text{solution}}}}. \label{pboundbig}
\end{equation}
This result can be simplified by multiplying the top and bottom by $\frac{L!}{\Omega^L} e^{\beta L\epsilon_{solution}}$,
resulting in
\begin{equation}
p_{bound}=\frac{e^{-\beta \epsilon_{\text{bound}}}
    \frac{\Omega^{L-1}}{(L-1)!} e^{-\beta (L-1)
    \epsilon_{\text{solution}}}}
    {\frac{\Omega^L}{L!} e^{-\beta
    L\epsilon_{\text{solution}}}
    +e^{-\beta \epsilon_{\text{bound}}}
    \frac{\Omega^{L-1}}{(L-1)!} e^{-\beta (L-1)
    \epsilon_{\text{solution}}}} \times
    \frac{\frac{L!}{\Omega^L} e^{\beta L\epsilon_{solution}}}
    {\frac{L!}{\Omega^L} e^{\beta L\epsilon_{solution}}}.
\end{equation}
We combine the two fractions in the previous equation and note that
$L!/(L-1)! = L$ and that $e^{-\beta \epsilon_{\text{bound}}}
e^{-\beta (L-1) \epsilon_{\text{solution}}} \times e^{-\beta L
\epsilon_{\text{solution}}} = e^{-\beta(\epsilon_{\text{bound}} -
\epsilon_{\text{solution}})}$.   Finally, we define the difference
in energy between a bound ligand and a ligand that is  free in
solution as $\Delta \epsilon = \epsilon_{\text{bound}} -
\epsilon_{\text{solution}}$. The probability of binding to the
receptor becomes
\begin{equation}
    p_{bound}=\frac{{L \over \Omega}e^{- \beta \Delta \epsilon}}{1+{L \over
    \Omega}e^{-\beta \Delta \epsilon}}.
\end{equation}
The overall volume of the box is $V_{\text{box}}$ and this permits
us to rewrite our results using concentration variables.  In
particular, this can be written in terms of ligand concentration
$c=L/V_{\text{box}}$ if we introduce $c_0=\Omega/V_{\text{box}}$, a
``reference'' concentration where every lattice position in the
solution is occupied. The choice of reference concentration is
arbitrary.   For the purposes of fig.~\ref{BindingCurve} we choose the elementary
box size to be 1~nm$^3$, resulting in $c_0 \approx 0.6$~M.  This is
comparable to the standard state used in
many biochemistry textbooks of 1~M.   The binding curve can
be rewritten as
\begin{equation}
    p_{\text{bound}}=\frac{{c \over c_0}e^{- \beta \Delta
    \epsilon}}{1+{c \over c_0}e^{- \beta \Delta \epsilon}}.
\label{pBoundLigandReceptorEquation}
\end{equation}
This classic result goes under many different names depending upon
the field.  In biochemistry, this might be referred to as a Hill
function with Hill coefficient one.  Chemists and physicists might
be more familiar with this result as the Langmuir adsorption
isotherm which provides a measure of the surface coverage as a
function of the partial pressure or the concentration. Regardless of
names, this expression will be our point of departure for thinking
about all binding problems and an example of this kind of binding
curve is shown in fig.~\ref{BindingCurve}.   This reasoning can be
applied to binding in real macromolecules of biological interest
such as myoglobin as shown in fig.~\ref{fig:MyoBound}.
 Though many problems of biological interest exhibit
binding curves that are ``sharper'' (i.e. there is a more rapid
change in $p_{bound}$ with ligand concentration) than this one,
ultimately, even those curves are measured against the standard
result derived here.

So far, we have examined binding from the perspective of statistical
mechanics.
That same problem can be addressed from the point of view of
equilibrium constants and the law of mass action, and it is
enlightening to examine the relation between the two points of view.
To see the connection,
the reaction of interest is characterized by the stoichiometric
equation \begin{equation} L+R \rightleftharpoons LR.
\end{equation}
This reaction is described by a dissociation constant given by the
law of mass action as
\begin{equation}\label{eq:eqMassAction}
    K_d={[L][R] \over [LR]}.
\end{equation}
It is convenient to rearrange this expression in terms of the
concentration of ligand-receptor complexes as \begin{equation}
[LR]={[L][R] \over K_d}. \label{LRKd}
\end{equation}
As before, our interest is in the probability that the receptor will
be occupied by a ligand. In terms of the concentration of free
receptors and ligand-receptor complexes, $p_{bound}$ can be written
as \begin{equation} p_{\text{bound}}={[LR] \over [R]+[LR]}.
\label{pboundLR2}
\end{equation}
We are now poised to write $p_{\text{bound}}$ itself by invoking
eq.~\ref{LRKd}, with the result that
\begin{equation}
    p_{\text{bound}} = \frac{ \frac{[L][R]}{K_d}}{[R] + \frac{[L][R]}{K_d}} =
        \frac{\frac{[L]}{K_d}}{1+\frac{[L]}{K_d}}. \label{pBoundThermo}
\end{equation}
What we see is that $K_d$ is naturally interpreted as that
concentration of ligand at which the receptor has a probability of
$1/2$ of being occupied.

We can relate our two descriptions of binding as embodied in
eqns.~\ref{pBoundLigandReceptorEquation} and ~\ref{pBoundThermo}.
Indeed, these two equations permit us to go back and forth between
the statistical mechanical and thermodynamic treatment of binding
through the recognition that the dissociation constant can be
written as $K_d=c_0e^{\beta \Delta \epsilon}$.   To see that, we
note that both of these equations have the same functional form
($p_{bound} = x/(1+x)$) allowing us to set $[L]/K_{d} =
\frac{c}{c_{0}} e^{-\beta \Delta \epsilon}$ and noting that $[L]=c$
by definition. This equation permits us to use measured equilibrium
constants to determine microscopic parameters such as the binding
energy as illustrated in figs.~\ref{BindingCurve} and
\ref{fig:MyoBound}.

\section{Statistical Mechanics and Transcriptional Regulation}
\label{Applications}


%





Transcriptional regulation is at the heart of much of biology. With
increasing regularity, the data that is being generated on
transcriptional regulation is quantitative.   In particular, it is
possible to quantify how much a given gene is expressed, where
within the organism and at what time.      Of course, with the
advent of such data, it is important that models of transcriptional
regulation keep pace with the measurements themselves.  The first
step in the transcription process is the binding of RNA polymerase
 to its target DNA sequence at the start of
a gene known as a promoter.   From a statistical mechanics
perspective, the so-called ``thermodynamic models'' of gene
expression are founded upon the assumption that the binding
probability of RNA polymerase can be used as a surrogate for the
extent of gene expression itself
\cite{Ackers1982,Buchler2003a,Bintu2005a,Bintu2005b}. Here too, the
use of equilibrium ideas must be justified as a result of separation
of time scales such that the binding step can equilibrate before the
process of transcription itself begins. The formulations derived
above can be directly applied to the process of transcription by
binding of RNA polymerase to DNA.   The fact that DNA is an
extended, linear polymer makes it seem at first blush like a very
different type of binding problem than the protein-ligand
interactions discussed above.    Nevertheless, as we will show
below, these same basic ideas are a natural starting point for the
analysis of gene expression.
%


%

\begin{figure}
\begin{center}
  \includegraphics[width=5in]{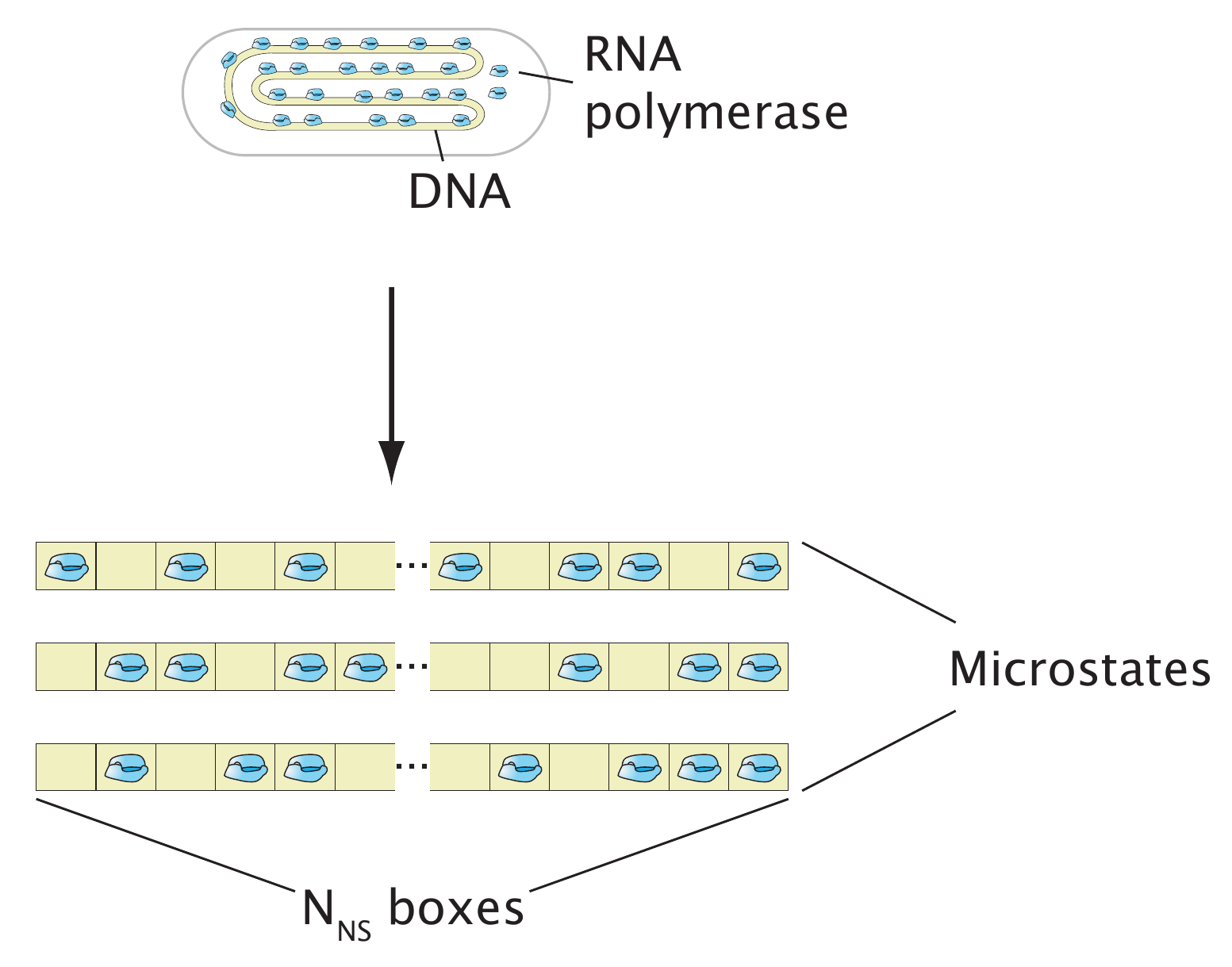}\\
  \caption{RNA polymerase nonspecific reservoir.  This figure represents DNA as a series of
  binding sites (schematized as boxes) for RNA polymerase.  The number of
  nonspecific binding sites is $N_{NS}$.  }
  \label{fig:RNAPReservoir}
\end{center}
\end{figure}

The way we set up the problem is shown in
fig.~\ref{fig:RNAPReservoir}. First, we argue that the genome can be
idealized as a ``reservoir'' of $N_{NS}$ nonspecific binding sites.
We assume that RNA polymerase may bind with its footprint starting
at any base pair in the entire genome, and that almost all of these
possible binding sites are nonspecific ones. For the case of {\it E.
coli}, there are roughly $500 - 2,000$ polymerase molecules for a
genome of around $5 \times 10^6$ base pairs \cite{Jishage1995}.
Amongst these nonspecific sites, there is one particular site (the
promoter for the gene of interest) that we are interested in
considering.  In particular, we want to know the probability that
this specific site will be occupied.

As noted above, the simplest model for RNA polymerase binding argues that the DNA
can be viewed as $N_{NS}$ distinct boxes where we need to place $P$
RNA polymerase molecules, only allowing one such molecule per site.
This results in the  partial partition function characterizing
the distribution of polymerase molecules on the nonspecific DNA as
\begin{equation}
Z_{NS}(P,N_{NS})=\underbrace{{N_{NS}! \over P! (N_{NS}-P)!}}_{\mbox{number of arrangements}} \times \underbrace{e^{-\beta P
\epsilon_{pd}^{NS}}}_{\mbox{Boltzmann weight}}.
\end{equation}
We will use the notation $\epsilon_{pd}^{S}$ to characterize the
binding energy of RNA polymerase to specific sites (promoters) and
$\epsilon_{pd}^{NS}$ to characterize the binding energy for
nonspecific sites.  (A note of caution is that this model is overly
simplistic since there is a continuum of different binding energies
for the nonspecific sites \cite{Gerland2002a}.) We are now poised to
write down the {\it total} partition function for this problem which
broadly involves two classes of states: i) all $P$ RNA polymerase
molecules are bound nonspecifically (note the similarity to the
partition function for ligand-receptor binding as shown in
eq.~\ref{eq:Zligandreceptor}), ii) one of the polymerase molecules
is bound to the promoter and the remaining $P-1$ polymerase
molecules are bound nonspecifically. Given these two classes of
states, we can write the \textit{total} partition function as
\begin{equation}
Z (P,N_{NS})    = \underbrace{Z_{NS}(P,N_{NS})}_{\mbox{empty
promoter}} + \underbrace{Z_{NS}(P-1,N_{NS}) e^{-\beta
\epsilon_{pd}^S}}_{\mbox{RNAP on promoter}}.
\end{equation}

\begin{figure}
\begin{center}
\includegraphics[width=5.3in]{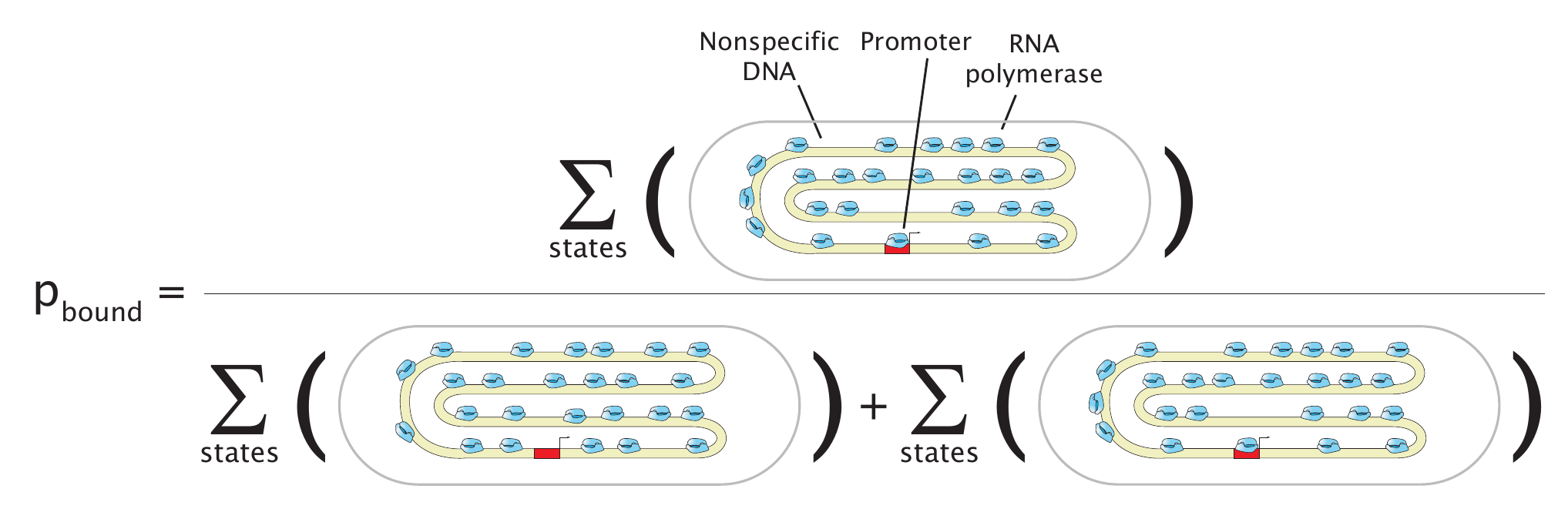}
\caption{Probability of RNA polymerase binding.  The probability of
polymerase binding is constructed as a ratio of favorable outcomes
(i.e. promoter occupied) to all possible outcomes. We are assuming that
RNA polymerase is always bound to DNA, either specifically or nonspecifically.} \label{Pbound}
\end{center}
\end{figure}

To find the probability that RNA polymerase is bound to the promoter
of interest, we compute the ratio of the weights of the
configurations for which the RNA polymerase is bound to its promoter
to the weights associated with all configurations. This is presented
schematically in fig.~\ref{Pbound} and results in
\begin{equation}
p_{bound}={{N_{NS}! \over (P-1)! (N_{NS}-(P-1))!} e^{-\beta (P-1) \epsilon_{pd}^{NS}} e^{-\beta
\epsilon_{pd}^S} \over
{N_{NS}! \over P! (N_{NS}-P)!}  e^{-\beta P \epsilon_{pd}^{NS}}
+ {N_{NS}! \over (P-1)! (N_{NS}-(P-1))!}  e^{-\beta (P-1) \epsilon_{pd}^{NS}} e^{-\beta
\epsilon_{pd}^S}}.
\label{pBoundPolymerase}
\end{equation}
Although this equation looks extremely grotesque it is really just
the same as eq.~\ref{pBoundLigandReceptorEquation} and is
illustrated in fig.~\ref{Pbound}.
   In order
to develop intuition for this result, we need to simplify the
equation by invoking the approximation $\frac{N_{NS}!}{(N_{NS}-P)!}
\simeq \left(N_{NS}\right)^{P}$, which holds if $P \ll N_{NS}$. Note
that this same approximation was invoked earlier in our treatment of
ligand-receptor binding. In light of this approximation, if we
multiply the top and bottom of eq.~\ref{pBoundPolymerase} by
$\left[P!/\left(N_{NS}^P\right)\right] e^{\beta P
\epsilon_{pd}^{NS}}$, we can write our final expression for
$p_{bound}$ as
\begin{equation}
p_{bound}(P,N_{NS}) = {1 \over {N_{NS} \over P} e^{\beta
(\epsilon_{pd}^S-\epsilon_{pd}^{NS})} +1 }.
\label{pbound1}
\end{equation}
Once again, it is the energy difference $\Delta \epsilon$ that
matters rather than the absolute value of any of the particular
binding energies. Furthermore the difference between a strong
promoter and a weak promoter can be considered as equivalent to a
difference in $\Delta \epsilon$ \cite{Bintu2005a}.

\begin{figure}
\begin{center}
\includegraphics[width=3.3in]{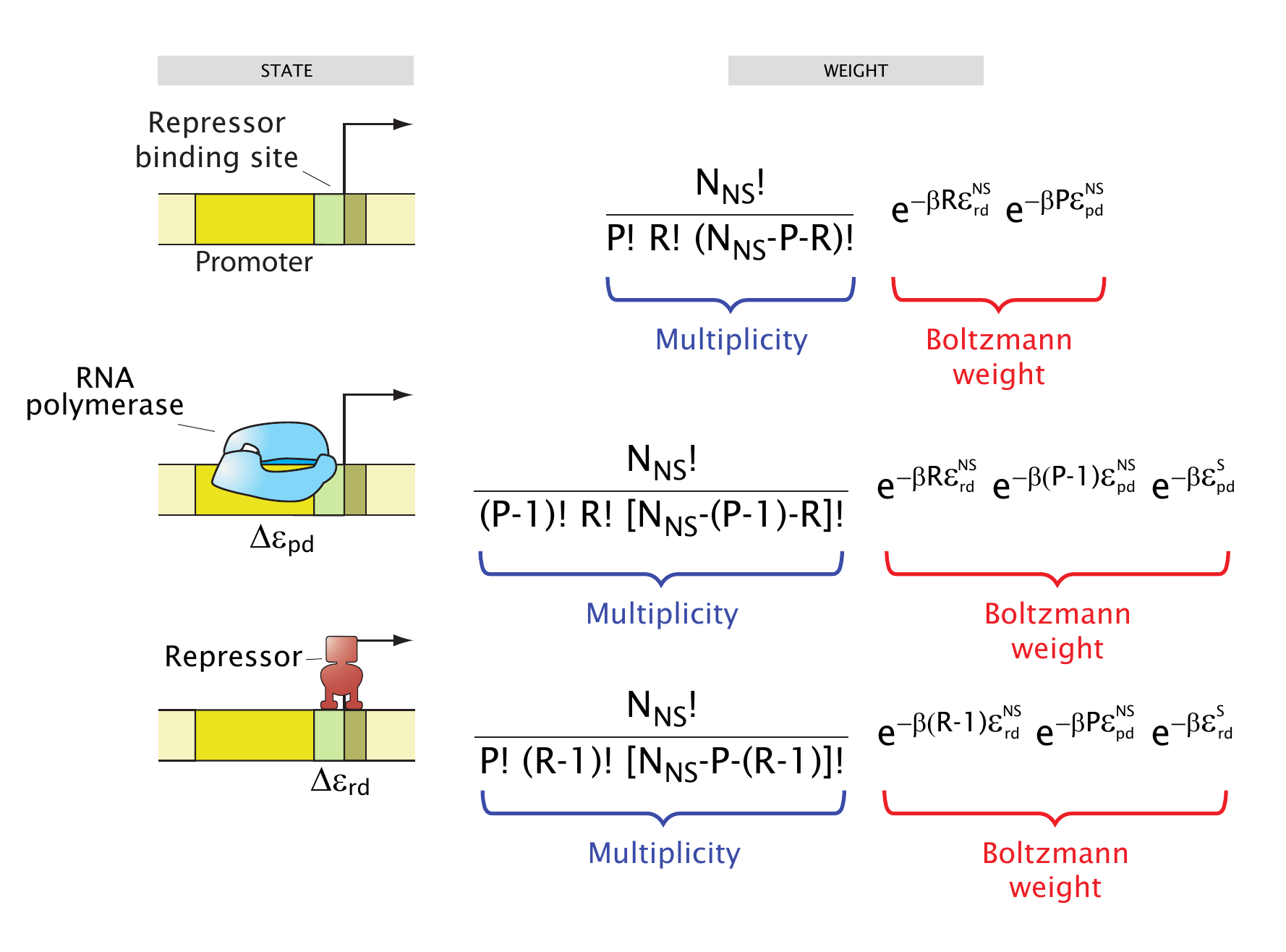}
\caption{States and weights for promoter in the presence of
repressor.   The promoter is labeled in dark yellow and the repressor
binding site (operator) is labeled in brown. Notice the overlap between the
promoter and the repressor binding site, which is denoted in green.   The weights of these
different states are a product of the multiplicity of the state of interest
and the corresponding Boltzmann factor. } \label{PromoterStatesWeights}
\end{center}
\end{figure}

The problem of promoter occupancy becomes much more interesting when
we acknowledge the fact that  transcription factors and other
proteins serve as molecular gatekeepers, letting RNA polymerase bind at
the appropriate time and keeping it away from the promoter at
others.       Different individual   transcription factors can
either activate the promoter through favorable molecular contacts
between the polymerase and the activator or they can stand in the
way of polymerase binding (this is the case of repressors). An
example of repression is shown in fig.~\ref{PromoterStatesWeights}.
The same ideas introduced above can be exploited for examining the
competition between repressor and polymerase for two overlapping
binding sites on the DNA. In this case, there are three classes of
states to consider: i) empty promoter, ii) promoter occupied by RNA
polymerase and iii) promoter occupied by repressor.    In this case,
the total partition function is given by
\begin{equation}
Z_{tot}(P,R,N_{NS}) = \underbrace{Z(P,R,N_{NS})}_{\mbox{empty promoter}} +
\underbrace{Z(P-1,R,N_{NS}) e^{-\beta \epsilon_{pd}^S}}_{\mbox{RNAP on promoter}}+
\underbrace{Z(P,R-1,N_{NS}) e^{-\beta \epsilon_{rd}^S}}_{\mbox{repressor on promoter}},
\end{equation}
where we have written the total partition function as a sum over
partial partition functions which involve sums over certain
restricted sets of states.  Each $Z$ is written as a function of $P$
and $R$, the number of polymerases and repressors in the nonspecific
reservoir, respectively. We have introduced $\epsilon_{rd}$, which
accounts for the energy of binding of the repressor to a specific
site ($\epsilon_{rd}^{S}$) or to a nonspecific site
($\epsilon_{rd}^{NS}$). For example, the term corresponding to the
empty promoter can be written as
\begin{equation}
    Z(P,R,N_{NS})=\frac{N_{NS}!}{P! R! (N_{NS} - P - R)!} \times
        e^{-\beta P \epsilon_{pd}^{NS}} \times e^{-\beta R \epsilon_{rd}^{NS}}.
\end{equation}
The other two terms have the same form except that $P$ goes to $P-1$
or that $R$ goes to $R-1$.

Once we have identified the various competing states and their weights, we
are in a position to ask key questions of biological interest.  For example,
what is the probability of promoter occupancy as a function of repressor concentration?
The results worked out above  now provide us with the tools in order to evaluate the probability that the promoter will
be occupied by RNA polymerase.  This probability is given by the ratio of the favorable outcomes to
all of the outcomes.  In mathematical terms, that is
\begin{equation}
p_{bound}(P,R,N_{NS})={Z(P-1,R,N_{NS}) e^{-\beta \epsilon_{pd}^S} \over
Z(P,R,N_{NS})+ Z(P-1,R,N_{NS}) e^{-\beta \epsilon_{pd}^S}+Z(P,R-1,N_{NS}) e^{-\beta \epsilon_{rd}^S}}.
\end{equation}
As argued above, this result can be rewritten in compact form by
dividing top and bottom by $Z(P-1,R,N_{NS}) e^{-\beta
\epsilon_{pd}^S}$ and by invoking the approximation
\begin{equation}
{N_{NS}! \over P! R! (N_{NS}-P-R)!} \simeq {\left(N_{NS}\right)^P \over P!}{\left(N_{NS}\right)^R \over R!}
\end{equation}
which amounts to the physical statement that there are so few
polymerase and repressor molecules in comparison with the number of
available sites, $N_{NS}$, that each of these molecules can more or
less fully explore those $N_{NS}$ sites without feeling the presence
of each other.
The resulting probability is
\begin{equation}\label{eq:pboundRep}
p_{bound}(P,R,N_{NS})   =   \frac{ 1}{ 1 + \frac{N_{NS}}{P} e^{\beta(\epsilon_{pd}^S-\epsilon_{pd}^{NS})}
                    (1 + {R \over N_{NS}} e^{-\beta (\epsilon_{rd}^S-\epsilon_{rd}^{NS})})}.
\end{equation}

%
%

\begin{figure}
\begin{center}
\includegraphics[width=3.3in]{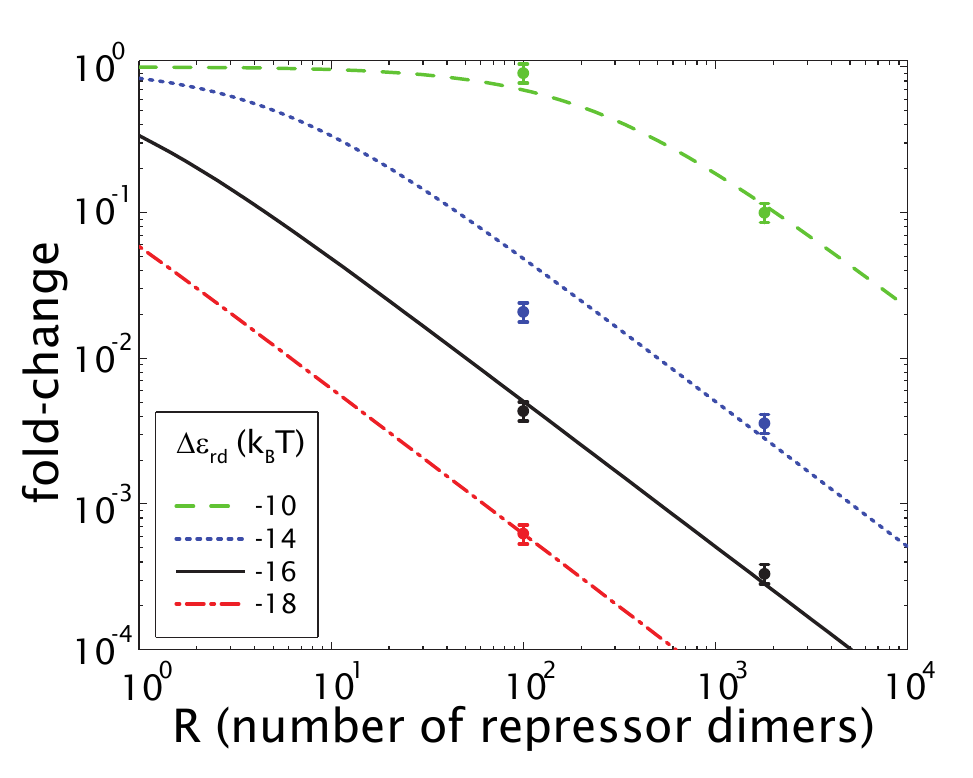}
\caption{Fold-change due to repression.  Experimental data on the fold-change in gene expression as
    a function of repressor concentration and binding affinity to DNA
    \cite{Oehler1994} and corresponding determination of the binding strength
    of repressor to DNA given by the theoretical model.  Experimentally, the
    different binding strengths are changed by
    varying the DNA sequence of the site where repressor binds. For each one of these DNA constructs
    only one parameter, the difference in binding energy $\Delta \epsilon$ is obtained using
    eq.~\ref{eq:foldchangeRepNoRNAP}.} \label{PboundPlot}
\end{center}
\end{figure}

Of course, a calculation like this is most interesting when it sheds
light on some experimental measurement.    In this case, we can
appeal to measurements on one of the classic regulatory networks in
biology, namely, the {\it lac} operon.  The {\it lac} promoter
controls genes that are responsible for lactose utilization by
bacteria. When the operon is ``on'', the bacterium produces the
enzymes necessary to import and digest lactose.  By way of contrast,
when the operon is ``off'', this enzymes are lacking (or expressed
at low ``basal'' levels).  It has been possible to measure relative
changes in the production of this enzyme as a function of both the
number of repressor molecules and the strength of the repressor
binding sites \cite{Oehler1994}. This relative change in gene
expression is defined as the concentration of protein product in the
presence of repressor divided by the concentration of protein
product in the absence of it. In order to connect this type of data
to the thermodynamic models we resort to one key assumption, namely
that the level of gene expression is linearly related to
$p_{\text{bound}}$, the probability of finding RNA polymerase bound
to the promoter. Once this assumption is made, we can compute the
fold-change in gene expression as
\begin{equation}\label{eq:foldchange}
    \text{fold-change in gene expression} =
    \frac{p_{\text{bound}}(R\neq0)}{p_{\text{bound}}(R=0)}.
\end{equation}
After inserting eq.~\ref{eq:pboundRep} in the expression for the
fold-change we find
\begin{equation}\label{eq:foldchangeRepRNAP}
    \text{fold-change in gene expression} =
    \frac{1+ \frac{N_{NS}}{P} e^{\beta (\epsilon_{pd}^S - \epsilon_{pd}^{NS})}}
    {1 + \frac{N_{NS}}{P} e^{\beta(\epsilon_{pd}^S-\epsilon_{pd}^{NS})}
                    (1 + {R \over N_{NS}} e^{-\beta (\epsilon_{rd}^S-\epsilon_{rd}^{NS})})}.
\end{equation}
Finally, we note that in the case of a weak promoter such as the
{\it lac} promoter, {\it in vitro} measurements suggest that the
factor $\frac{N_{NS}}{P}
e^{\beta(\epsilon_{pd}^S-\epsilon_{pd}^{NS})}$ is of the order of
500 \cite{Bintu2005a}. This makes the second term in the numerator
and denominator of eq.~\ref{eq:foldchangeRepRNAP} much bigger than
one. In this particular case of weak promoter the fold-change
becomes independent of RNA polymerase as follows
\begin{equation}\label{eq:foldchangeRepNoRNAP}
    \text{fold-change in gene expression} \simeq
    \frac{\frac{N_{NS}}{P} e^{\beta (\epsilon_{pd}^S - \epsilon_{pd}^{NS})}}
    {\frac{N_{NS}}{P} e^{\beta(\epsilon_{pd}^S-\epsilon_{pd}^{NS})}
                    (1 + {R \over N_{NS}} e^{-\beta (\epsilon_{rd}^S-\epsilon_{rd}^{NS})})} =
    \left( 1+ \frac{R}{N_{NS}} e^{-\beta (\epsilon_{rd}^S-\epsilon_{rd}^{NS})}\right)^{-1}.
\end{equation}
For binding to a strong promoter, this approximation is no longer
valid since the factor   $\frac{N_{NS}}{P}
e^{\beta(\epsilon_{pd}^S-\epsilon_{pd}^{NS})}$ is of  order 3 in
this case. In fig.~\ref{PboundPlot} we show the experimental data
for different concentrations of repressor and values of its binding
relative affinity to DNA, $\Delta \epsilon_{rd}=\epsilon_{rd}^{S} -
\epsilon_{rd}^{NS}$. Overlaid with this data we plot the fold-change
in gene expression given by eq.~\ref{eq:foldchangeRepNoRNAP} where
the binding strength to each type of site is determined. Notice that
for each curve there is only one parameter to be determined, $\Delta
\epsilon_{rd}$. This simple example shows how statistical mechanics
arguments can be used to interpret measurements on gene expression
in bacteria.

%


%


%

\section{Cooperativity in Binding: The Case Study of Hemoglobin}

So far, our treatment of binding has focused on simple binding
reactions such as the myoglobin binding curve in
fig.~\ref{fig:MyoBound} which do not exhibit the sharpness seen in
some biological examples. This sharpness in binding is often denoted
as ``cooperativity'' and refers to the fact that in cases where
several ligands bind simultaneously, the energy of binding is not
additive. In particular, cooperativity refers to the fact that the
binding energy for a given ligand depends upon the number of ligands
that are already bound to the receptor. Intuitively, the
cooperativity idea results from the fact that when a ligand binds to
a protein, it will induce some conformational change.  As a result,
when the next ligand binds, it finds an altered protein interface
and hence experiences a different binding energy (characterized by a
different equilibrium constant). From the point of view of
statistical mechanics, we will interpret cooperativity as an
interaction energy - that is, the energy of the various ligand
binding reaction are not simply additive.

 The classic example of this
phenomenon is hemoglobin, the molecule responsible for carrying
oxygen in the blood. This molecule has four distinct binding sites,
reflecting its structure as a tetramer of four separate
myoglobin-like polypeptide chains. Our treatment of ligand-receptor
binding  in the
 case of hemoglobin can be couched in the language of
two-state occupation variables.  In particular, for hemoglobin, we
describe the state of the system with the four variables
$(\sigma_1,\sigma_2,\sigma_3, \sigma_4)$, where $\sigma_i$ adopts
the values $0$ (unbound) or $1$ (bound) characterizing the occupancy
of site $i$ within the molecule.  One of the main goals of a model
like this is to address questions such as the average number of
bound oxygen molecules as a function of the oxygen concentration (or
partial pressure).

\begin{figure}
\begin{center}
\includegraphics[width=3.0in]{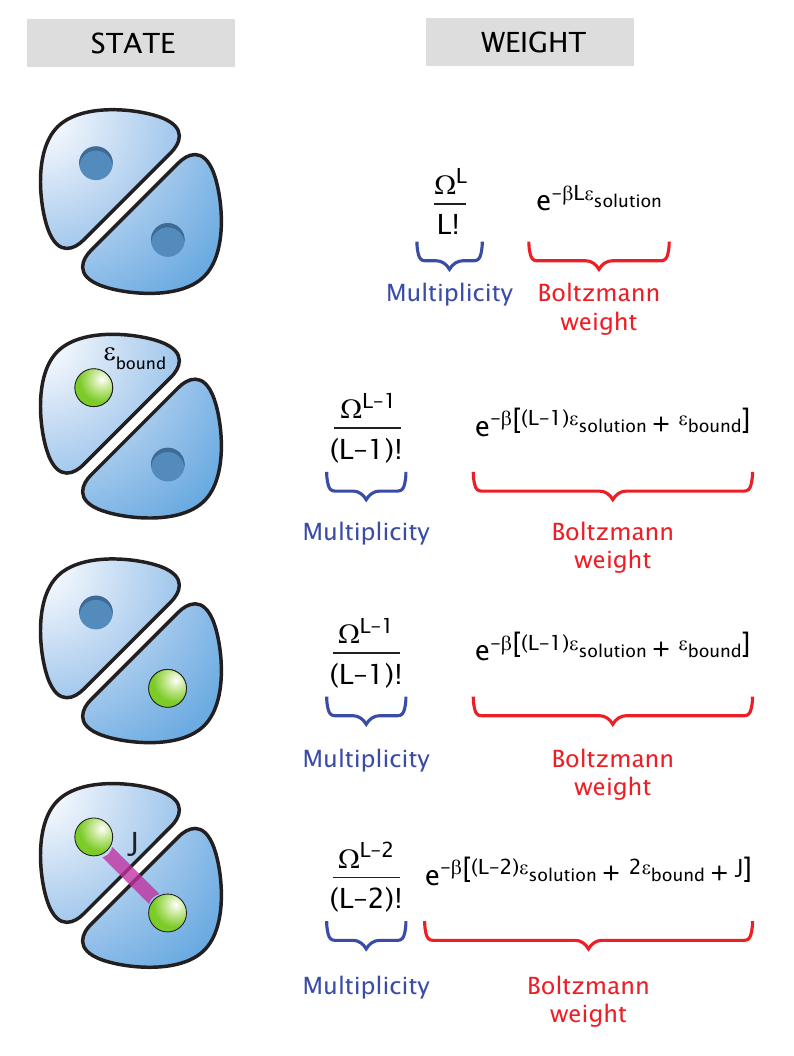}
\caption{States and weights for dimoglobin.  The different states correspond
to different occupancies of the two binding sites by oxygen molecules.}
\label{DimoglobinStateWeights}
\end{center}
\end{figure}

As a first foray into the problem of cooperative binding, we examine
a toy model which reflects some of the full complexity of binding in
hemoglobin.  In particular, we imagine a fictitious dimoglobin
molecule which has only two myoglobin-like polypeptide chains and
therefore two $O_{2}$ binding sites. We begin by identifying the
states and weights as shown in fig.~\ref{DimoglobinStateWeights}.
This molecule is characterized by four distinct states corresponding
to each of the binding sites of the dimoglobin molecule either being
occupied or empty.  For example, if binding site $1$ is occupied we
have $\sigma_1=1$ and if unoccupied then $\sigma_1=0$.  The energy
of the system can be written as
\begin{equation}
E=\Delta \epsilon(\sigma_1+\sigma_2)+J \sigma_1 \sigma_2,
\end{equation}
where $\Delta \epsilon$ is the energy gain garnered by the molecules
when  bound to dimoglobin as opposed to wandering around in
solution.  The parameter $J$ is a measure of the cooperativity and
implies that when both sites are occupied, the energy is more than
the sum of the individual binding energies.

\begin{figure}
\begin{center}
\includegraphics[width=3.0in]{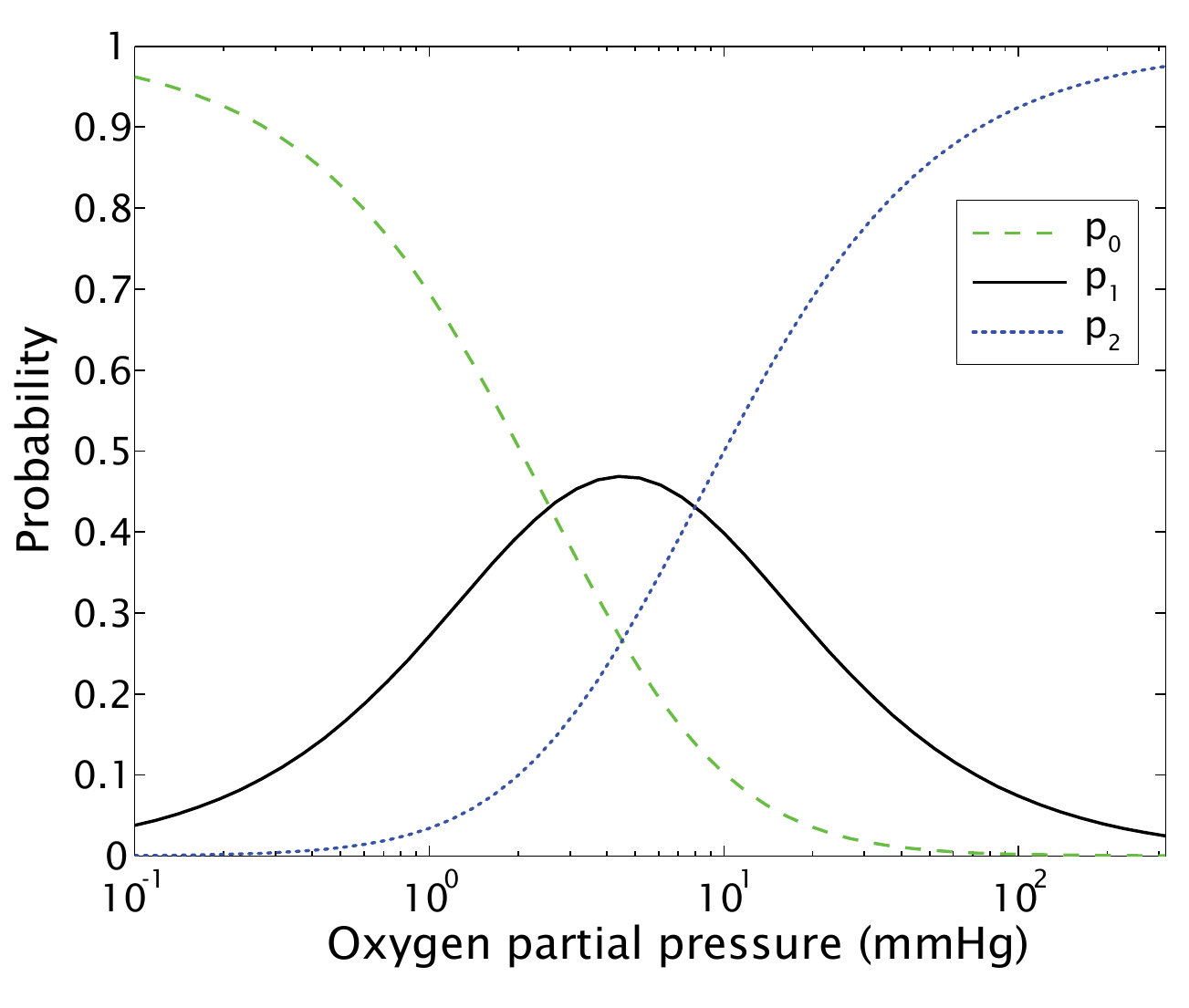}
\caption{Probabilities of oxygen binding to dimoglobin. The plot shows the probability of
finding no oxygen molecules bound to dimoglobing ($p_0$), of finding one molecule bound ($p_1$) and
that of finding two molecules bound ($p_2$). The parameters used are $\Delta\epsilon = -5~k_BT$, $J=-.25~k_BT$,
and $c_{0} = 760~\text{mmHg}$.}
\label{fig:DimoglobinProbabilities}
\end{center}
\end{figure}

As we have done throughout the article, we can compute the
probability of different states using the states and weights diagram
by constructing a ratio with the numerator given by the weight of
the state of interest and the denominator by the sum over all
states. In analogy to eq.~\ref{eq:ZLigandReceptor} we can write the
partition function corresponding to the weights in
fig.~\ref{DimoglobinStateWeights}. Next we can calculate the
probability of finding no oxygen molecule bound to our dimoglobin
molecule, the probability of finding one oxygen molecule bound or
that of finding two molecules bound. For example, if we want to
compute the probability of single occupancy we can add up the
weights corresponding to this outcome of
fig.~\ref{DimoglobinStateWeights} and divide them by the total
partition function which yields
\begin{equation}
    p_{1} = \frac{2 \frac{\Omega^{L-1}}{(L-1)!} e^{-\beta \left[ (L-1) \epsilon_{\text{solution}}
                + \epsilon_{\text{bound}}\right]}}{\frac{\Omega^L}{L!} e^{-\beta L \epsilon_{\text{solution}}} +
            2 \frac{\Omega^{L-1}}{(L-1)!} e^{-\beta \left[ (L-1) \epsilon_{\text{solution}}
                + \epsilon_{\text{bound}}\right]} +
            \frac{\Omega^{L-2}}{(L-2)!} e^{-\beta \left[ (L-2) \epsilon_{\text{solution}}+
                2\epsilon_{\text{bound}} + J \right]}}.
\end{equation}
Similarly to what was done in eq.~\ref{pBoundLigandReceptorEquation}
we can write the previous probability in terms of the standard state
and multiply and divide by $\frac{\Omega^L}{L!} e^{-\beta L
\epsilon_{\text{solution}}}$. This results in
\begin{equation}
    p_{1} = \frac{2 \frac{c}{c_0} e^{-\beta \Delta \epsilon}}
            {1 +
            2 \frac{c}{c_0} e^{-\beta \Delta \epsilon} +
            \left(\frac{c}{c_0}\right)^{2} e^{-\beta \Delta \epsilon + J }}.
\end{equation}
In fig.~\ref{fig:DimoglobinProbabilities} we plot this probability
as a function of the oxygen partial pressure as well as $p_{0}$ and
$p_{2}$, the probabilities of the dimoglobin molecule being empty
and being occupied by two oxygen molecules, respectively.

Next, we calculate the average number of bound oxygen molecules as a
function of its partial pressure. We add the number of molecules
bound in each state times the probability of that state occurring
\begin{equation}
    \langle N_{\text{bound}} \rangle = 1 \times \,p_{1} + 2 \times \,p_{2}=\frac{2
        \frac{c}{c_0} e^{-\beta \Delta \epsilon}+2 (\frac{c}{c_0})^2
        e^{-\beta(2\Delta \epsilon+J)}}{1+2\frac{c}{c_0} e^{-\beta \Delta
        \epsilon}+ (\frac{c}{c_0})^2 e^{-\beta(2\Delta \epsilon+J)}}.
\end{equation}
To further probe the nature of cooperativity, a useful exercise is
to examine the occupancy of the dimoglobin molecule in the case
where the interaction term $J$ is zero. We find that the average
occupancy is given by the sum of two independent single-site
occupancies as
\begin{equation}
\langle N \rangle = \frac{2 \frac{c}{c_0} e^{-\beta \Delta \epsilon}+2 (\frac{c}{c_0})^2
        e^{-\beta 2\Delta \epsilon}}
        {1+2\frac{c}{c_0} e^{-\beta \Delta
        \epsilon}+ (\frac{c}{c_0})^2 e^{-\beta 2\Delta \epsilon}}=
        \frac{2 \frac{c}{c_0} e^{-\beta \Delta \epsilon} \left(1+ \frac{c}{c_0}
        e^{-\beta \Delta \epsilon}\right)}
        {\left( 1+\frac{c}{c_0} e^{-\beta \Delta
        \epsilon}\right)^2} =
        2 {{c \over c_0} e^{-\beta \Delta \epsilon} \over 1+ {c \over c_0}e^{-\beta \Delta \epsilon}}.
\end{equation}



\begin{figure}
\begin{center}
\includegraphics[width=4.0in]{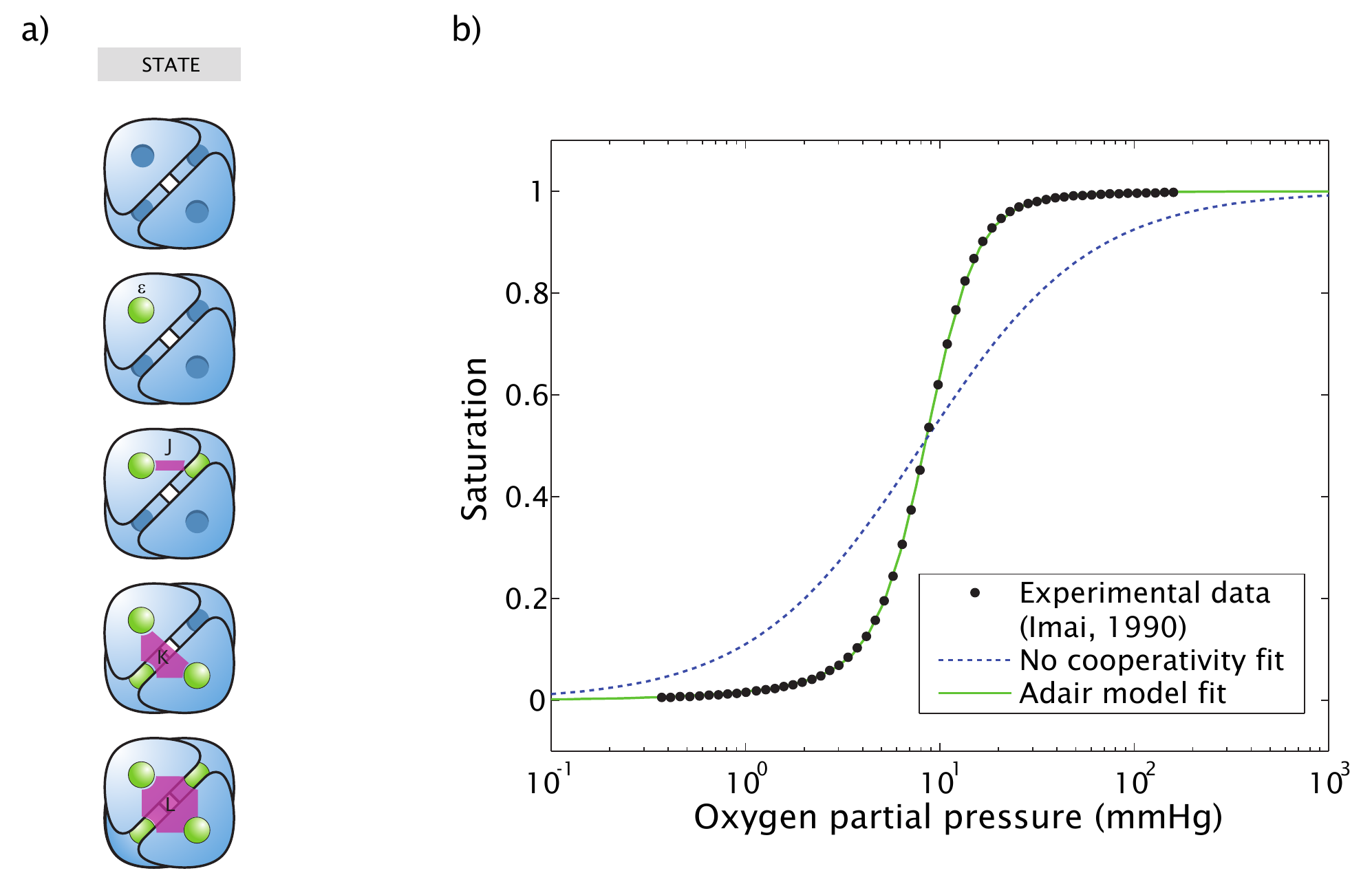}
\caption{Binding of oxygen by hemoglobin. (a) States considered in the Adair model and (b) plot of
    experimental data \cite{Imai1990} together with the
    data fit to the four parameters of the Adair model and to a simple no-cooperativity case.
    Note that the shape of the curve without cooperativity  looks different from that in fig.~\ref{fig:MyoBound} only
    because we plot it here using a log-scale.} \label{HemoglobinStatesWeights}
\end{center}
\end{figure}

 In considering the real
hemoglobin molecule rather than the fictitious dimoglobin, the only
novelty incurred is extra mathematical baggage.   In this case,
there are four state variables $\sigma_1$, $\sigma_2$, $\sigma_3$
and $\sigma_4$ that correspond to the state of oxygen occupancy at
the four distinct sites on the hemoglobin molecule.   There are
various models that have been set forth for thinking about binding
in hemoglobin, many of which can be couched simply in the language
of these occupation variables. One important model which was
introduced by Adair in 1925 \cite{Adair1925} assigns distinct interaction
energies for the binding of the second, third and fourth
oxygen molecules. The energy in this model is written as
\begin{equation}
E=\Delta \epsilon \sum_{\alpha=1}^4 \sigma_{\alpha}+ J\sum_{(\alpha \ne
\gamma)=1}^4  \sigma_{\alpha}\sigma_{\gamma}+K\sum_{(\alpha \ne \beta \ne \gamma) = 1}^{4}
\sigma_{\alpha}\sigma_{\beta}\sigma_{\gamma}+
L\sum_{(\alpha \ne \beta \ne \gamma \ne \delta)=1}^{4} \sigma_{\alpha}\sigma_{\beta}\sigma_{\gamma} \sigma_{\delta},
\end{equation}
where the parameters $K$ and $L$ capture the energy of the three-
and four-body interactions, respectively. This model results in a
very sharp response compared to a simple binding curve without any
cooperativity. In fig.~\ref{HemoglobinStatesWeights} we show the
corresponding states for this model as well as the fit to the
experimental data. This curve should be contrasted with the fit
curve using the simple binding model such as
eq.~\ref{pBoundLigandReceptorEquation}.

\section{Conclusions}

We have argued that statistical mechanics needs to be part of the
standard toolkit of biologists who wish to understand the
biochemical underpinnings of their discipline.   Indeed, the ideas
in this paper  represent a small part of a more general text on
physical biology entitled ``Physical Biology of the Cell'' worked on
by all of us which aims not only to provide the quantitative
underpinnings offered by statistical mechanics, but a range of other
tools that are useful for the quantitative analysis of living
matter. Our own experiments in teaching such material convince us
that a first exposure to statistical mechanics can be built around a
careful introduction of the concept of a microstate and the
assertion of the Boltzmann distribution as the fundamental ``law''
of statistical mechanics. These formal ideas in conjunction with
simple, approximate models such as the lattice model of solution and
two-state models for molecular conformations permit an analysis of a
large number of different interesting problems.

We challenge the notion that biologists need to understand every detail of
statistical mechanics in order to use it fruitfully in their thinking and research.
The importance of this dictum was stated eloquently by Schawlow who
noted ``To do successful research, you don't need to know everything.  You just
need to know of one thing that isn't known.''    To successfully apply
statistical mechanics, we argue that a feeling for microstates and how
to assign them probabilities will go a long way toward demystifying
statistical mechanics and permit biology students to think about many new problems.

With the foundations described in this paper in hand, our courses
turn to a variety of other interesting applications of statistical
mechanics that include: accessibility of nucleosomal DNA,
force-extension curves for DNA, lattice models of protein folding,
the origins of the Hill coefficient, the role of tethering effects
in biochemistry (what we like to call ``biochemistry on a leash''),
the Poisson-Boltzmann equation, the analysis of polymerization and
molecular motors and many more.   Though it is easy to make blanket
statements about living systems being far from equilibrium, the
calculus of equilibrium as embodied in statistical mechanics still
turns out to be an exceedingly useful tool for the study of living
systems and the molecules that make them work.


\section*{Appendix: A derivation of the Boltzman distribution} \label{app:Boltzmann}

\begin{figure}
\begin{center}
\includegraphics[width=3.0truein]{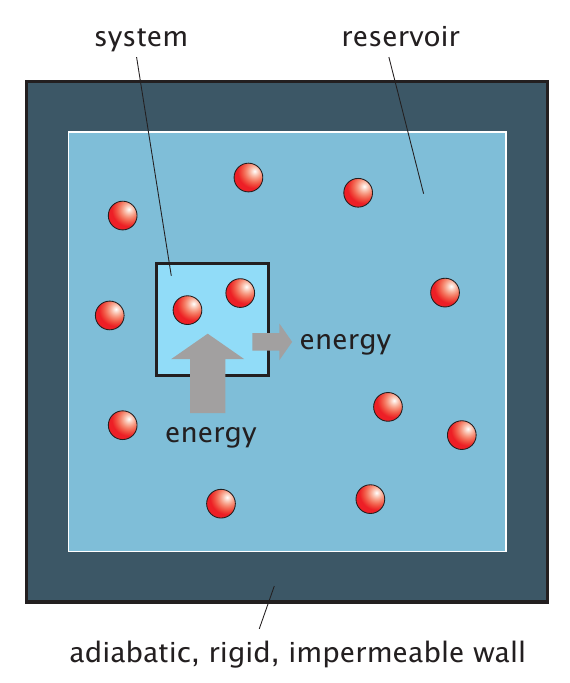}
\caption[]{System in contact
with a heat bath (thermal reservoir).  The system and its reservoir are completely
isolated from the rest of the world by walls that are adiabatic (forbid the
flow of heat out or in), rigid and impermeable (forbid the flow of matter).  Energy
can flow across the wall separating the system from the reservoir and as a result,
the energy of the system (and reservoir) fluctuate.}
\label{BoltzmannDistributionCartoon}
\end{center}
\end{figure}

The setup we consider for our derivation of the probability of
microstates for systems in contact with a  thermal reservoir is
shown in fig.~\ref{BoltzmannDistributionCartoon}. The idea is that
we have a box which is separated from the rest of the world by
rigid, impermeable and adiabatic walls.   As a result, the total
energy and the total number of particles within the box are
constant.   Inside this box, we now consider two regions, one that
is our {\it system} of interest and the other of which is the
reservoir. We are interested in how the system and the reservoir
share their energy.

The total energy is $E_{tot}=E_r+E_s$ where
the subscripts $r$ and $s$ signify reservoir and system, respectively.
Our fundamental assertion is that the probability of finding a given state of the system $E_s$
is proportional to the number of states available to the  {\it reservoir} when the system is
in this state.  That is
\begin{equation}
{p_s^{(1)} \over p_s^{(2)} } = {W_r(E_{tot}-E_s^{(1)})
\over W_r(E_{tot}-E_s^{(2)}) },
\label{BoltzmannProbs}
\end{equation}
where $W_r(E_{tot}-E_s^{(1)}) $ is the number of states available
to the {\it reservoir}, when the system is in the particular state $E_s^{(1)}$.  By constructing
the ratio of the probabilities, we avoid ever having to compute the absolute
number of states available to the system.
The logic is that we assert that the system is in {\it one} particular microstate that
is characterized by its energy $E_s$.  When the system is assigned this energy,
the reservoir has available a particular number of states  $W_r(E_{tot}-E_s)$
which depends upon how much energy, $E_{tot}-E_s$, it has.
Though the equations may seem cumbersome, in fact, it is the underlying conceptual idea
that is the subtle (and beautiful) part of the argument.  The point is that the total number of
states available to the universe of system plus reservoir when the system is in the
particular state $E_s^{(1)}$  is given by
\begin{equation}
W_{tot}(E_{tot}-E_s^{(1)}) = \underbrace{1}_{\mbox{states of system}}
\times \underbrace{W_r(E_{tot}-E_s^{(1)}) }_{\mbox{states of reservoir}}
\end{equation}
because we have asserted that the system itself is in one particular
microstate which has energy $E_s^{(1)}$.  Though there may be other
microstates with the same energy, we have selected one particular
microstate of that energy  and ask for its probability.

We now ask what the relative probability between two states is.
Basically, instead of counting the number of microstates available
to a particular states we calculate the relative difference in
microstates available to two given states (1) and (2). This will
allow us to calculate the probabilities of each macrostate up to a
multiplicative factor. As we saw in the text, this multiplicative
factor will be given by the partition function $Z$. We can then
rewrite eq.~\ref{BoltzmannProbs} as
\begin{equation}
 {W_r(E_{tot}-E_s^{(1)})
\over W_r(E_{tot}-E_s^{(2)}) }= {e^{S_r(E_{tot}-E_s^{(1)})/k_B }
\over e^{S_r(E_{tot}-E_s^{(2)})/k_B } },
\end{equation}
where we have invoked the familiar Boltzmann equation for the entropy, namely $S=k_B \mbox{ln} ~W$ which
can be rewritten as $W =e^{S/k_B}$.
To complete the derivation, we now note that $E_s << E_{tot}$.
As a result, we can expand the entropy as
\begin{equation}
S_r(E_{tot}-E_s) \approx S_r(E_{tot}) -{\partial S_r \over \partial E} (E_{tot}-E_s),
\end{equation}
where we have only kept terms that are first order in the differences.
Finally, if we recall the thermodynamic identity $(\partial S / \partial E) =1/T$,
we can write our result as
\begin{equation}
{p_s^{(1)} \over p_s^{(2)} } ={e^{-E_s^{(1)}/k_BT }
\over e^{-E_s^{(2)}/k_BT }}
\end{equation}
The resulting probability for finding the system in  state $i$  with energy $E_s^{(i)}$ is
\begin{equation}
p_s^{(i)}={e^{-\beta E_s^{(i)}} \over Z},
\end{equation}
precisely the Boltzmann  distribution introduced earlier.

\section*{Acknowledgments}

We are extremely grateful to a number of people who have given us
both guidance and amusement in thinking about these problems (and
some  for commenting on the manuscript):   Phil Nelson,
Dan Herschlag, Ken Dill, Kings Ghosh, Mandar Inamdar.
JK acknowledges the
support of NSF  DMR-0403997 and is a Cottrell Scholar of Research
Corporation. RP acknowledges the support of the NSF  and the NIH
Director's Pioneer Award.
 HG is grateful for support from both the
NSF funded NIRT and the NIH Director's Pioneer Award. JT is
supported by the NIH.

\bibliographystyle{pnas}
\bibliography{PaperLibrary,AJPPapers}

\end{document}